\documentclass[11pt,a4paper]{article}
\usepackage{jheppub}
\usepackage{amsfonts}

\usepackage{epsfig}
\usepackage{float}
\usepackage{latexsym}
\usepackage{graphicx}
\usepackage{amssymb}
\usepackage{amsmath}
\usepackage{color}

\title{Regular Phantom Black Hole and Holography: very high temperature superconductors}
\author[a]{Kai Lin,}
\author[a]{A. B. Pavan,}
\author[b]{Qiyuan Pan,}
\author[b]{and E. Abdalla}

\affiliation[a]{Universidade Federal de Itajub\'a, Instituto de F\'isica e Qu\'{\i}mica, \\
CEP 37500-903, Itajub\'a, Brazil}

\affiliation[b]{Instituto de F\'isica, Universidade de S\~ao Paulo,\\
CEP 05315-970, S\~ao Paulo, Brazil}

\emailAdd{lk314159@hotmail.com}
\emailAdd{alan@unifei.edu.br}
\emailAdd{panqiyuan@126.com}
\emailAdd{eabdalla@usp.br}

\date{\today}

\abstract{
 Holographic superconductors containing a non-minimal derivative coupling for scalar field in a regular phantom plane symmetric black hole have been considered. We show that the parameter of the regular black hole $b$ as well as the non-minimal derivative coupling parameter $\eta$ affect the formation of the condensate as well as the conductivity in the superconductor. Moreover, $b$ has a critical value in which the critical temperature $T_c$ increases without a bound. We argue that an unlimited critical temperature is an evidence that high $T_c$ superconductor must be related to the absence of a singularity in the bulk in the AdS/CFT context.}


\begin{document}

\maketitle

\flushbottom
 \newcommand{\bq}{\begin{equation}}
 \newcommand{\eq}{\end{equation}}
 \newcommand{\bqn}{\begin{eqnarray}}
 \newcommand{\eqn}{\end{eqnarray}}
 \newcommand{\nb}{\nonumber}
 \newcommand{\lb}{\label}

\section*{Introduction}

Black holes are gravitational objects whose gravity is so overwhelming that not even light can escape their realm. In Einstein gravity, such solutions do exist with varying dynamics  admitting just three measurable properties: mass, charge and angular momentum, the so-called black hole's hairs. Therefore, despite  their exotic condition, black holes are simple gravitational objects in theoretical physics. Because of its simplicity it is a ideal research object for quantum gravity.

An important region of black holes is their singularity at the origin, which leads to a failure of the physical
description at that region. Thus, the proposition of the cosmic censorship hypothesis requiring that
the singularity should be covered by a horizon. Nevertheless, the singularity (even if it is not visible from future null infinity)
is still a trouble for modern physics.

In order to solve the singularity problem, Bardeen
proposed a new idea that introduces some
geometrical parameters into the metric of the black hole
\cite{Regular0} such that they could cancel the divergence of
physical quantities at the singular point. This new kind of object is
named \emph{regular black hole}. According to this idea, the
singularity problem could be resolved by geometrical methods
without introducing the quantum gravity hypothesis. This simple
method has attracted a lot of interest. Some recent results show
that a non-linear electromagnetic field could be a candidate to provide
regularity  \cite{RegularCharge,RegularChargeA,RegularChargeB}. However, the complete understanding of the non-linear electromagnetic field is still an
open problem. Thus, one looks for a regular neutral black hole solution \cite{RegularNonCharge,RegularNonChargeA,RegularNonChargeB,RegularNonChargeC,RegularNonChargeD,RegularNonChargeE,RegularNonChargeF} or use another kind of sources \cite{Regularother,RegularotherA,RegularotherB,RegularotherC}.

In 2006, Bronnikov and Fabris considered a phantom scalar field evolving in a spherically symmetric spacetime in the context of the Einstein gravity and found a regular de Sitter black hole solution where the phantom field can cancel the singularity \cite{Regular1}. Due to the fact that the phantom field is a candidate for dark energy, this phantom regular black hole implies that the singularity of black hole  could be avoided by dark energy.

On the other hand, in 2009, Hartnoll, Herzog and Horowitz built a
holographic description for a superconductor by coupling a scalar field with the electromagnetic field evolving in Schwarzschild-AdS spacetime, and used the AdS/CFT (Anti-de Sitter/Conformal Field Theory) correspondence to explain how this configuration was interpreted as a holographic superconductor model \cite{holographic0,holographic0A}. Such a work reveals some relations between
gravitational theory, quantum field theory and condensed matter
physics, and provided a new idea to investigate condensed matter physics. Subsequently, the research of holographic superconductors and their phase transitions in several black hole spacetimes with higher and lower dimensions and different couplings between fields has been performed \cite{holographic1,holographic1A,holographic1B,holographic1C,holographic1D,holographic1E,holographic1F,holographic1G,holographic1H,holographic1I,holographic1J,holographic1L,holographic1M,holographic1N,holographic1O,holographic1P,holographic1Q,holographic1U,holographic1V,holographic1X}. Thus it would be interesting to know how a regular black hole affects the properties of a holographic superconductor.

In this paper we studied the holographic superconductor with non-minimal derivative coupling scalar field in a regular
phantom plane-symmetric black hole.

In order to use the AdS/CFT correspondence appropriately we first derive the metric of regular phantom plane-symmetric AdS black
hole and prove that the singularity inside the horizon disappears. In section III, a holographic superconductor model
with non-minimal derivative coupling scalar field is constructed. In section IV some numerical results for phase transition curves and the electrical conductivity are obtained for the model. The last section includes some discussion about the critical value of $b$ with fixed $\eta$ and other conclusions.

\section{Metric of Regular Planar Anti-de Sitter Phantom Black Hole}

Recent research and observations of cosmology show that our universe
is expanding in an accelerated way, and the mysterious object driving
the accelerated expansion is named dark energy. Many models for dark energy can be used to correct the Einstein field equation and
several candidates of dark energy are proposed, such as cosmological constant,
quintessence, phantom and k-essence. As an important candidate, the
phantom field possesses some strange property, and would be very
interesting to consider the effect of phantom in black hole
spacetime. In 2006, the authors of \cite{Regular1} obtained an exact phantom
regular spherically symmetric de Sitter black hole solution, and in this section we
follow their idea to derive the phantom regular plane-symmetric AdS black hole metric.

We consider the Lagrangian of a self-gravitating scalar field $\phi$ with an
arbitrary potential $V(\phi)$ and the metric of static planar black
hole,
 \bq
\label{eq1} {\cal
L}=R-\epsilon\nabla^\alpha\phi\nabla_\alpha\phi-2V(\phi),
 \eq
and the corresponding plane-symmetric Ansatz for the metric,
 \bq
\label{eq2} ds^2=-f(r)dt^2+\frac{dr^2}{f(r)}+p(r)^2(dx^2+dy^2),
 \eq
where $\epsilon =1$ corresponds to the normal scalar field while phantom
field requires $\epsilon =-1$. We choose $\epsilon=-1$  in this paper. Moreover, we set $p(r)^2$ instead of
$r^2$ in front of $dx^2+dy^2$ term because the regulator parameter $b$
is  included in $p(r)$. In the solution of Bronnikov
and Fabris, $p(r)=\sqrt{r^2+b^2}$, and the black hole reduces to
Schwarzschild-de Sitter black hole as $b\to 0$.

In this paper, we  consider $p(r)=\sqrt{r^2+b^2}$ in static plane-symmetric spacetime. Subsequently, substituting Eq.(\ref{eq2}) into the scalar field and Einstein equations, we get
 \bqn
\label{eq3}
\frac{d}{dr}\left(fp^2\frac{d\phi}{dr}\right)+p^2\frac{dV}{d\phi}&=&0\quad ,\\
\frac{d}{dr}\left(p^2\frac{df}{dr}\right)+2p^2V&=&0\quad ,\\
2\frac{d^2p}{dr^2}-\left(\frac{d\phi}{dr}\right)^2&=&0\quad ,\\
f\frac{d^2p^2}{dr^2}-p^2\frac{d^2f}{dr^2}&=&0\quad .
 \eqn
This set of equations can be exactly solved resulting in the solution
\bqn
\label{eq4}
f(r)&=&(r^2+b^2)\left\{\frac{c}{b^2}+\frac{r_0}{b^3}\left[\frac{br}{r^2+b^2}+\arctan\left(\frac{r}{b}\right)\right]\right\}\quad ,\\
\label{eq4a}
\phi(r)&=&-\sqrt{2}\arctan\left(\frac{r}{b}\right)+\phi_0\quad ,\\
\label{eq4b}
V(\phi)&=&\frac{1}{2\sqrt{2}b^3}\left\{4r_0\phi-3\sqrt{2}r_0\sin\left(\sqrt{2}(\phi_0-\phi)\right)\right.\nb\\
&&\left.-2\cos\left(\sqrt{2}(\phi_0-\phi)\right)\left[r_0\phi-\sqrt{2}cb-r_0\phi_0\right]\right\}\quad ,
 \eqn
 where $c$, $b$, $r_0$ and $\phi_0$ are undetermined constants. The solution $f(r)$  in (\ref{eq4}) can have different interpretations just like the spherically symmetric solution obtained by \cite{Regular1}. At this moment, we are interested in the range of the constants where the solution (\ref{eq4}) behaves like an asymptotic AdS black hole. The other interpretations will be discussed in a future work. Following the prescription used in \cite{Regular1}, $r_0$ can be related to the mass of the black hole and $c$ is responsible to change the asymptotic behaviour of $f(r)$. For the sake of convenience, in this paper, we choose to work with $r_e$ instead of the mass of the black hole. Thus, we will set
 \bqn
 c&=&b^2-\frac{\pi r_0}{2b},\\
 \nb\\
 r_0&=&\frac{b^3}{\frac{\pi}{2}-\arctan\left(\frac{r_e}{b}\right)-\frac{br_e}{(r_e^2+b^2)}},
 \eqn
so that $f$ behaves as $r^2$ as $r\rightarrow\infty$, but vanishes at the event horizon $r_e$. Under these considerations the components of the metric, the scalar field potential and the scalar field become
\bqn
\label{eq4c}
f(r)&=&\frac{(r^2+b^2)\left[h(r)-h(r_e)\right]}{\frac{\pi}{2}-h(r_e)},\ \textrm{where},\ h(r)=\frac{br}{r^2+b^2}+\arctan\left(\frac{r}{b}\right)\, \\
\nb\\
\label{eq4d}
\phi(r)&=&-\sqrt{2}\arctan\left(\frac{r}{b}\right)+\phi_0\, \\
\nb\\
\label{eq4e}
V(\phi)&=&\frac{1}{2\sqrt{2}b^3}\left\{4r_0\phi-3\sqrt{2}r_0\sin\left(\sqrt{2}(\phi_0-\phi)\right)\right.\nb\\
&&\left.-2\cos\left(\sqrt{2}(\phi_0-\phi)\right)\left[r_0\phi-\sqrt{2}cb-r_0\phi_0\right]\right\}\quad .
\eqn

\begin{figure*}[h]
\includegraphics[width=8.5cm]{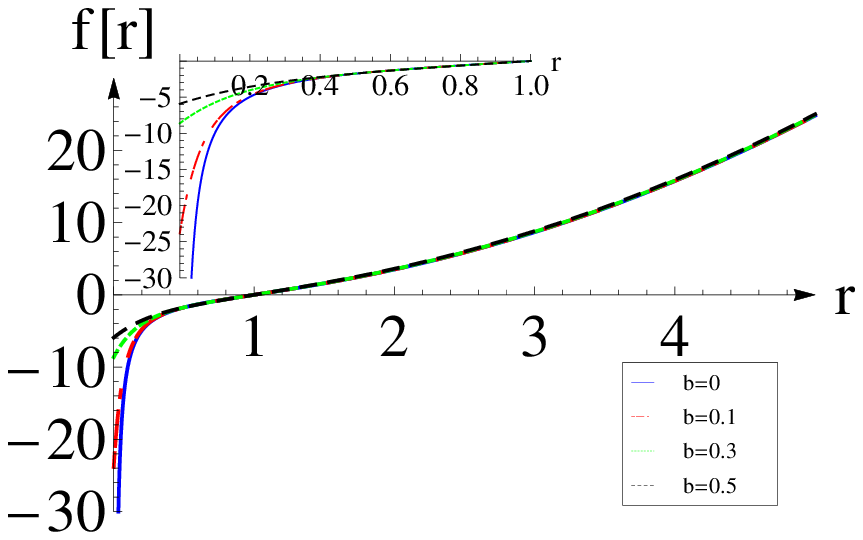}~~~\includegraphics[width=8.5cm]{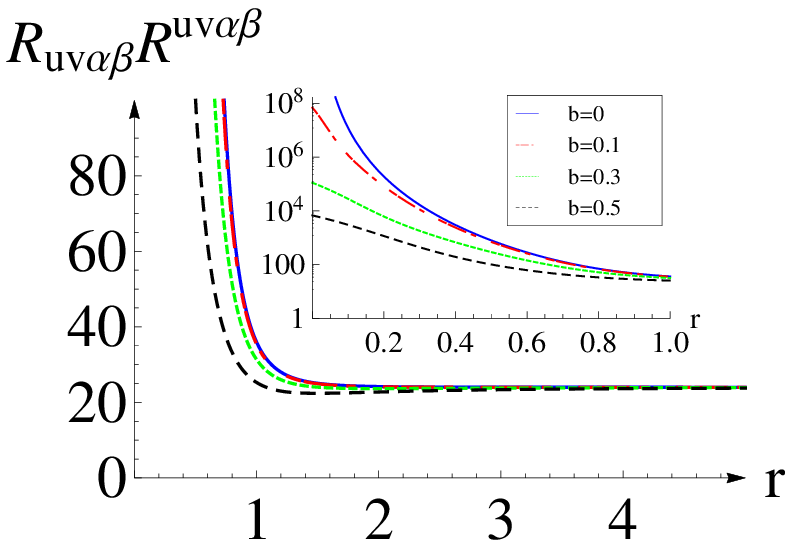}\\
\caption{The function $f(r)$ (left) and the Kretschmann scalar (right) for different values of $b$ with $r_e=1$.}
\label{figN1}
\end{figure*}

\begin{figure*}[h]
\includegraphics[width=8cm]{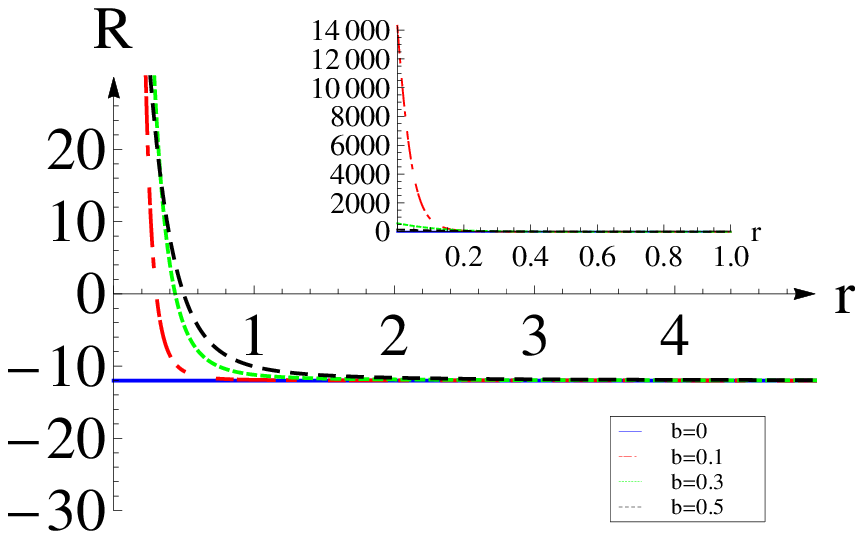}~~~\includegraphics[width=8cm]{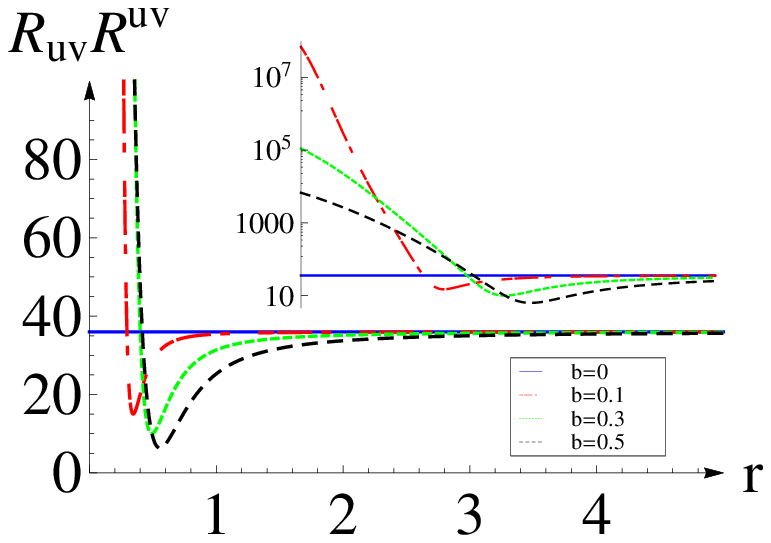}
\caption{The Ricci scalar $R$ (left) and the invariant scalar $R_{\mu\nu}R^{\mu\nu}$ (right) for different values of $b$ with $r_e=1$.}
\label{figN2}
\end{figure*}

In order to compare the regular planar black hole with the planar Schwarzschild AdS black hole (setting $b=0$), we have plotted the function $f(r)$ and the scalars of curvature, $R$, $R_{\mu\nu}R^{\mu\nu}$ and $R_{\mu\nu\alpha\beta}R^{\mu\nu\alpha\beta}$ in Figs.(\ref{figN1},\ref{figN2}). Inspecting the results we can see that the behaviour for $f(r)$ and the scalars of curvature in the region far from the event horizon is quite similar for all black hole solutions. It means that the regularity at the origin of black hole does not affect its structure in the asymptotic spatial infinity. This special feature is important because we want to apply the AdS/CFT dictionary in the holographic superconductor context using this regular black hole solution as a gravitational background.

On the other hand, in the region inside of the event horizon $r<r_{e}$, both $f(r)$ and the Kretschmann scalar diverge at $r=0$ for the planar Schwarzschild AdS black hole whereas for the regular phantom planar black hole they stay finite. The Ricci scalar and the invariant scalar $R_{\mu\nu}R^{\mu\nu}$ are always finite for all black hole solutions but their behaviour near the origin depends on $b$. They decrease when $b$ increases. We also observe that  a minimum value of the invariant scalar $R_{\mu\nu}R^{\mu\nu}$ exists and it
changes with $b$.

The Hawking temperature of a planar black hole is obtained computing $T_h =\frac{f'(r_e)}{4\pi}$  resulting in
 \bqn
\label{eq5}
T_h=\frac{b^3}{\pi}\left\{(b^2+r_e^2)\left[\pi-2\arctan\left(\frac{r_e}{b}\right)\right]-2br_e\right\}^{-1}\quad .
 \eqn
From  Eq.(\ref{eq5}) one can see that $b\geq0$ for a non-negative temperature.

In order to compare the Hawking temperature between the planar Schwarzschild AdS black hole and the regular black hole we expand the Eq.(\ref{eq5}) around $b=0$, that results
\bqn
\label{eq5a}
T_h&\approx & \frac{3r_e}{4\pi}+\frac{3b^2}{20\pi
r_e}+{\cal O}(b^4),\\
\nb
T_h&\approx & T_h^{SchAdS}\left(1+\frac{b^2}{5 r_e^2}\right)+{\cal O}\left(b^4\right).
 \eqn
For the same event horizon $r_e$ the temperature of the regular black hole $T_h$ is always larger than the temperature of Schwarzschild AdS $T_h^{SchAdS}$. If one repeats the same analysis for the component of the metric we find
\bqn
f(r)\approx \left(r^2-\frac{r_h^3}{r}\right)+\left(1-\frac{6 r_h}{5 r}+\frac{r_h^3}{5 r^3}\right) b^2+{\cal O}(b^4)\quad .
 \eqn
Thus, one can see that the regularity on the origin has a strong influence around the event horizon affecting the Hawking temperature and metric component.
Summarizing the results we can see that the above equations show that the metric of the regular black hole looks like the planar Schwarzschild black hole metric with small corrections in the limit of $b\to 0$. In view of the AdS/CFT dictionary we argue that the regularity on the origin could be interpreted as fluctuations in the temperature in quantum field theory on the boundary increasing its value. It will increase the value of the critical temperature of the holographic superconductor rendering the condensation easier than in a planar Schwarzschild AdS.

In the next section we shall construct a holographic superconductor model with non-minimal coupling in the derivative of the scalar field using the regular planar black hole.

\section{Field equations of holographic superconductor with a scalar
field coupled to the Einstein tensor}

In order to generalize the holographic superconductor model we will introduce an additional non-minimal coupling between the charged scalar field and spacetime curvature besides those for the Abelian field $A_\mu$. Some possible non-minimal couplings are $R\partial_\mu\Psi\partial^\mu\Psi$, $R^{\mu\nu}\partial_\mu\Psi\partial_\nu\Psi$ and $R\Psi\nabla^2\Psi$. Of course, the coupling with a Einstein tensor $G^{\mu\nu}\partial_\mu\Psi\partial_\nu\Psi$ is the most common representative form \cite{nonminimalmais,nonminimalmaisA,nonminimalmaisB,nonminimalmaisC,nonminimalmaisD,nonminimalmaisF}. Moreover, in the context of cosmology, a non-minimally coupled scalar field was used as a candidates for dark energy and dark matter by several authors \cite{nonminimal,nonminimalA,nonminimalB,nonminimalC,nonminimalD,nonminimalE}. Therefore, it is meaningful to probe the influence of the \emph{dark sector} together with the phantom scalar field in condensed matter devices through such a non-minimal coupling .
Here, we consider a holographic superconductor model with a non-minimal coupling $G^{\mu\nu}\partial_\mu\Psi\partial_\nu\Psi$, with the action

 \bqn
\label{eq6} {\cal L}_{HS}=-\frac{1}{4}F_{uv}F^{uv}-\left(g^{uv}+\eta
G^{uv}\right)\nabla_u\Psi\nabla_v\Psi-q^2\Psi^2A_uA^u+\frac{2}{L_\eta^2}\Psi^2\quad ,
 \eqn

where $\eta$ is the  non-minimal coupling constant, $q$ and $L_\eta^2$ are the charge and a constant related to the mass of the scalar field, respectively. We will work in the probe limit in which $A_\mu$ and $\Psi$ do not backreact on the metric. Considering the probe fields as initially static we derive two coupled equations of motion from the above action and the metric, respectively, resulting in

 \bqn
\label{eq7psi}
\Psi''&+&\mathcal{P}\ \Psi'+\mathcal{Q}\ \Psi=0\quad,\\
\nb\\
\label{eq7phi}
\Phi''&+&2\frac{p'}{p}\Phi'-\frac{2q^2\Psi^2}{f}\Phi=0\quad ,\\
\nb\\
\mathcal{P}&=& \frac{\eta  f p' \left(3 f'p' +2 f p''\right)+p^2 f'}{f\left(\eta p f' p' +\eta fp'^2+p^2\right)}+\frac{p \left[\eta f'^2 p' +f \left(p' \left(\eta f''+2\right)+\eta f' p'' \right)\right]}{f\left(\eta p f' p' +\eta fp'^2+p^2\right)}\quad,\\
\nb\\
\mathcal{Q}&=&\frac{p^2 \left(2 f+L_\eta^2 q^2 \Phi^2\right)}{L_\eta^2f^2 \left(\eta  p f' p' +\eta f p'^2+p^2\right)}\quad ,
\eqn

where $A_\mu=\Phi\delta^0_\mu$. Without loss of generality, we set $q=1$ and the position of the event horizon is chosen as $r_e=1$. The asymptotic behaviour of the fields at  infinity is given by
 \bqn
\label{eq8psi}
\Psi&=&\frac{A_1(\eta)}{ r^{\Delta_+}}+\frac{B_1(\eta)}{ r^{\Delta_-}}\cdots\quad ,\\
\nonumber\\
\label{eq8delta}
\Delta_{\pm}&=&\frac{3}{2}\mp\frac{3}{2}\sqrt{1-\frac{8}{9 L_\eta^2(1+3\eta)}}\quad ,\\
\nonumber\\
\label{eq8phi}
\Phi&=&A_2+\frac{B_2}{r}+\cdots\quad .
 \eqn
For simplicity and convenience we set $L_\eta^2=(1+3\eta)^{-1}$ so that the conformal dimension of the operators are $\Delta_+=1$ and $\Delta_-=2$ independently of the value of $\eta$. Thus, the influence of the non-minimal coupling will only appear explicitly in the equation of motion for the scalar field Eq.(\ref{eq7psi}). The properties of dual field theory are related to the asymptotic behaviour of the fields. The expectation-value of the scalar operators $\mathcal{O}_{1,2}$ in the field theory dual to the field $\Psi$ can be read off the Eq.(\ref{eq8psi}),
\bqn
\label{eq8a}
\langle\mathcal{O}_1\rangle=\sqrt{2}A_1, \ \ \ \ \langle\mathcal{O}_2\rangle=\sqrt{2}B_1,
 \eqn
whereas the chemical potential $\mu$ and the charge density $\rho$ in the field theory dual to the field $\phi$ is obtained in Eq.(\ref{eq8phi}). Therefore, the asymptotic behaviour of the fields can be written in terms of the quantities of the theory on the boundary,
\bqn
\label{eq8b}
\Psi&=&\frac{\langle\mathcal{O}_1\rangle}{\sqrt{2}r} + \frac{\langle\mathcal{O}_2\rangle}{\sqrt{2}r^2} \cdots \,,\\
\nonumber\\
\label{eq8c}
\Phi&=&\mu-\frac{\rho}{r}+\cdots\; .
 \eqn
In this case, $\Psi$ has two normalizable modes, so one can impose the appropriate boundary conditions to calculate the diagrams of phase transition for the holographic superconductor.
\begin{figure*}
\includegraphics[width=5.5cm]{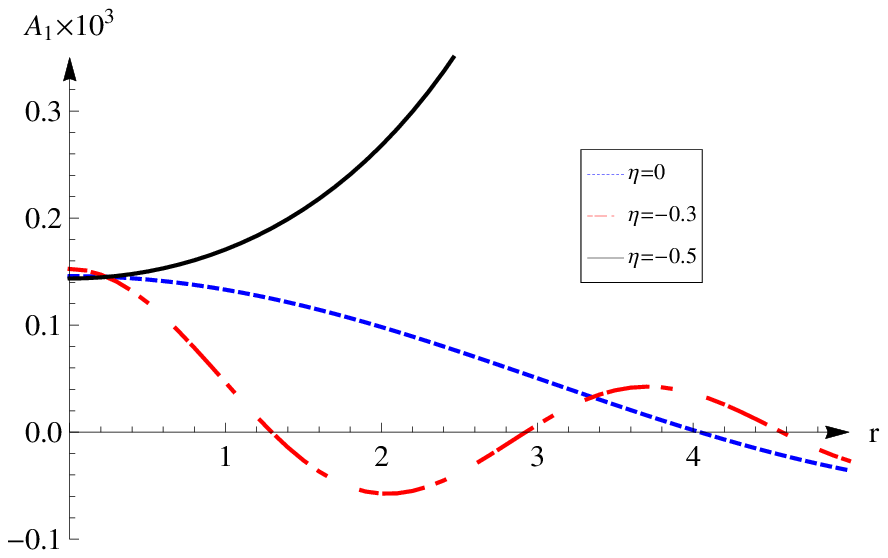}~~~\includegraphics[width=5.5cm]{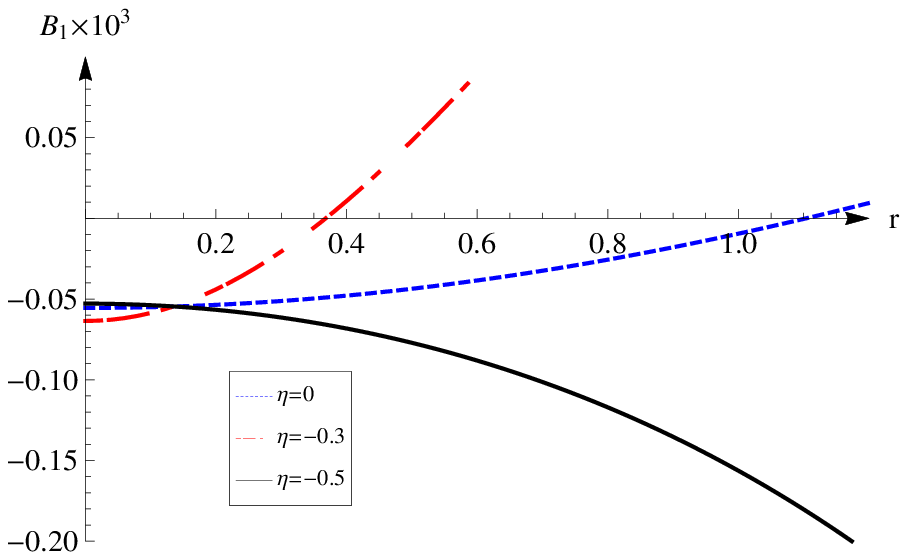}
\caption{The behaviour of asymptotic scalar field $\Psi$ around to $\eta=0$ with $b=0.1$}
\label{fig0}
 \end{figure*}
However, as  discussed above, $\Psi$ depends implicitly on $\eta$ affecting its asymptotic behaviour. In order to have normalizable modes described by  Eq.(\ref{eq8b}) we found, numerically, that $\eta$ must be larger than $-1/3$. This can be seen in Fig.(\ref{fig0}).

\section{Diagrams of phase transitons}

Considering the boundary conditions at infinity and at the event horizon respectively, we use the shooting method to calculate the diagrams of phase transitions numerically.
\begin{figure*}[h]
\includegraphics[width=4cm]{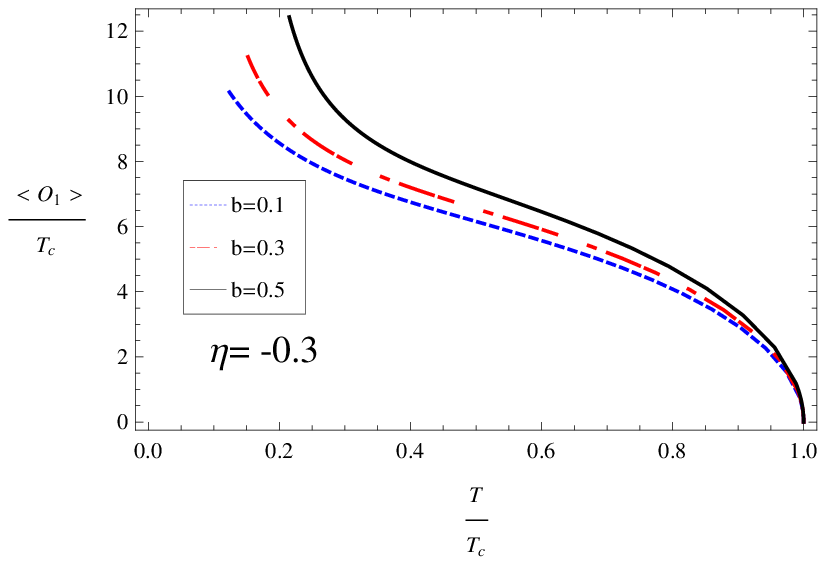}\includegraphics[width=4cm]{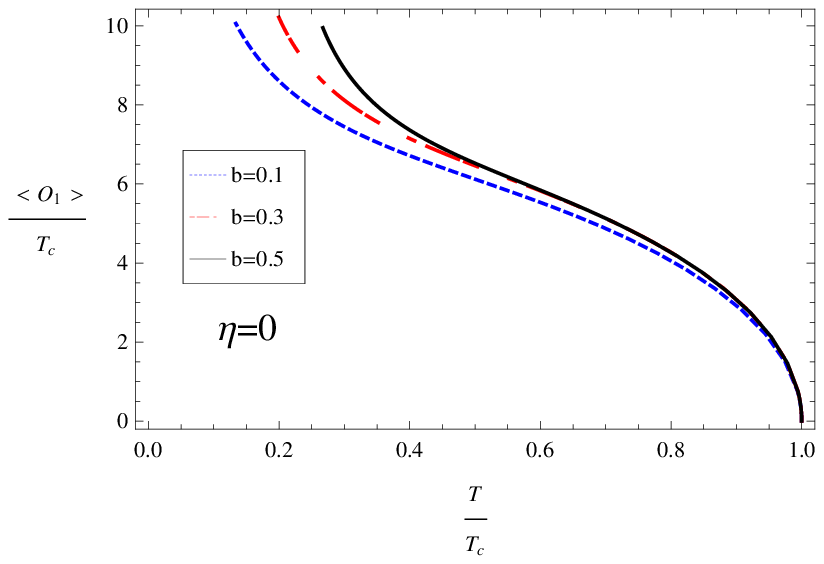}\includegraphics[width=4cm]{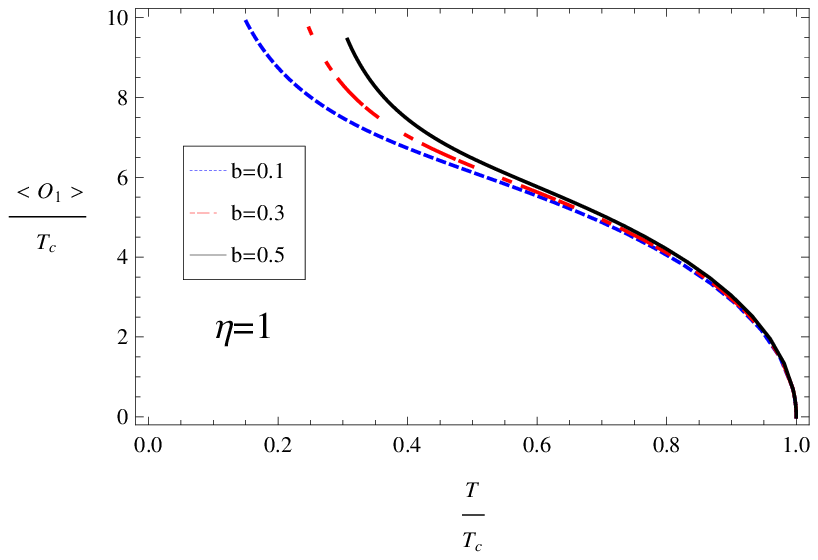}\includegraphics[width=4cm]{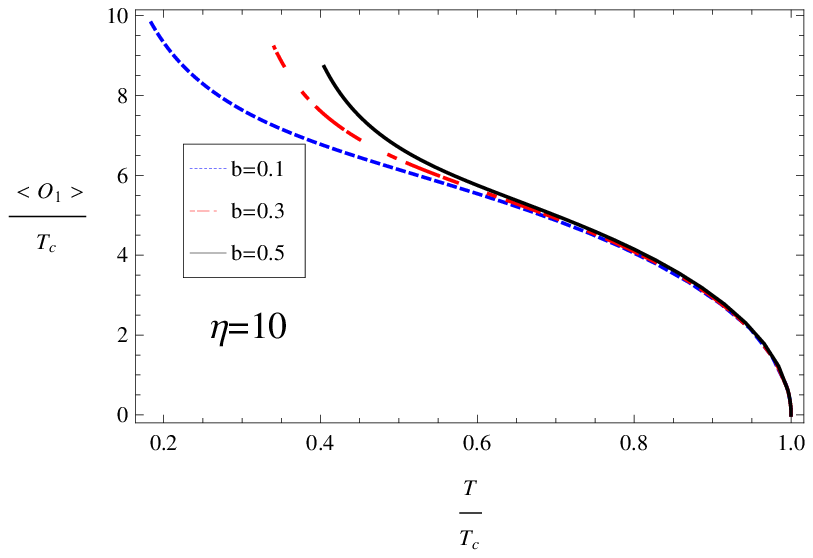}
\includegraphics[width=4cm]{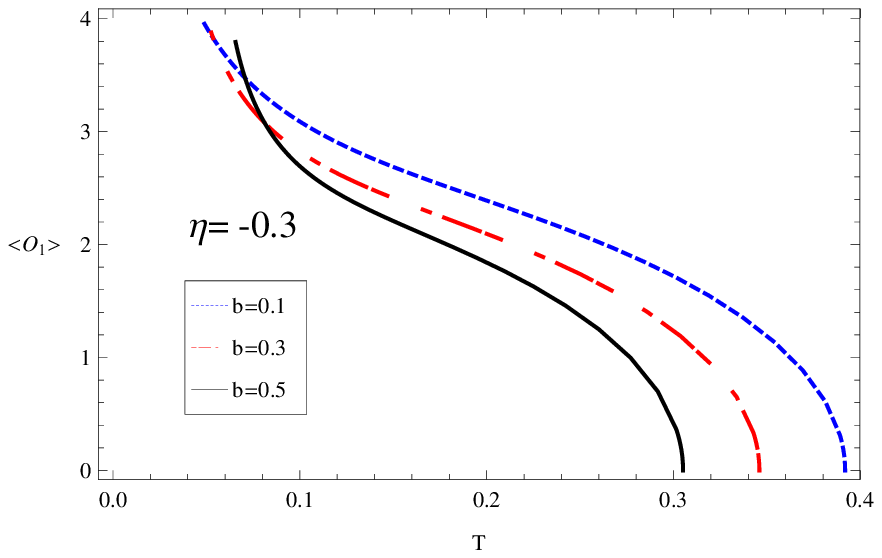}\includegraphics[width=4cm]{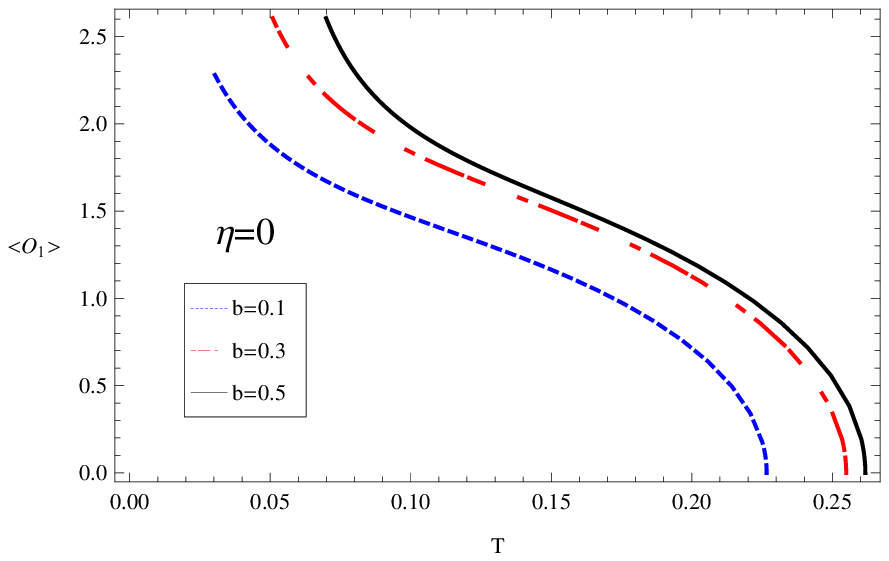}\includegraphics[width=4cm]{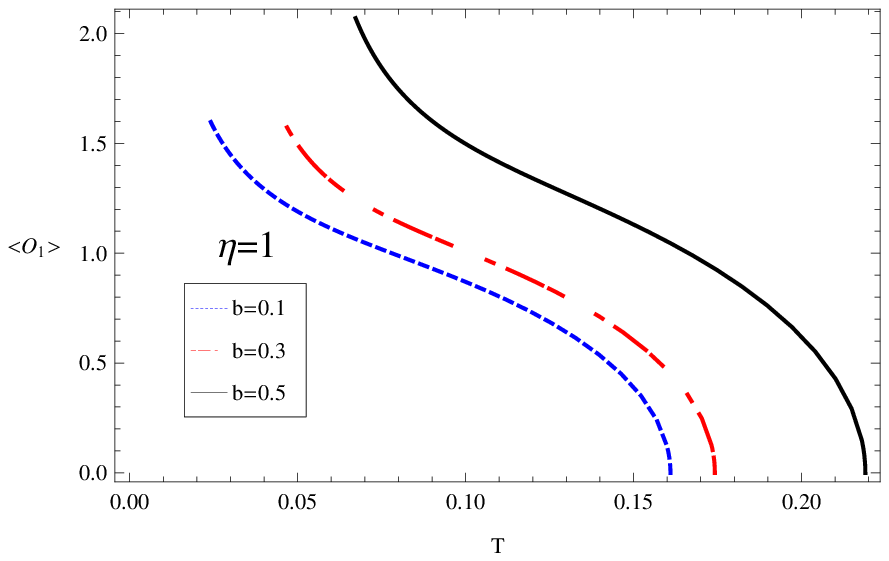}\includegraphics[width=4cm]{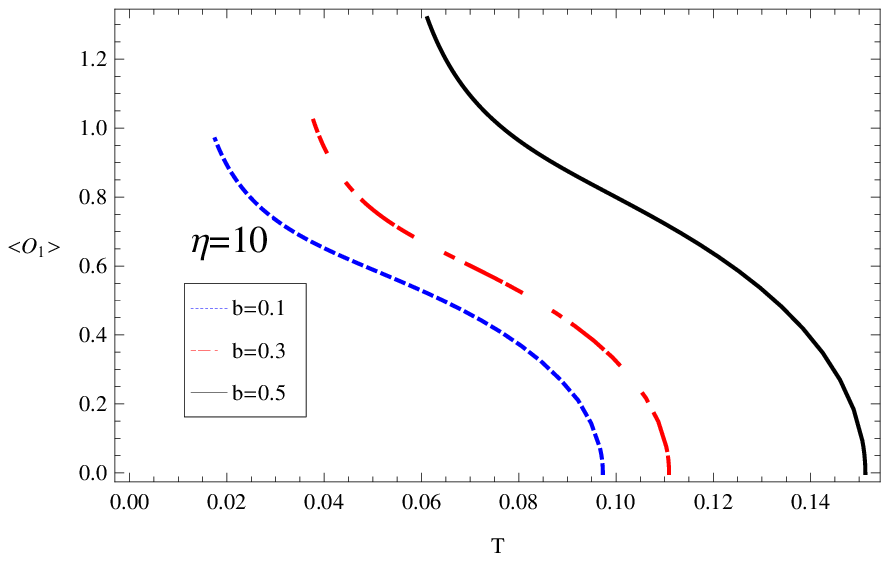}
\includegraphics[width=4cm]{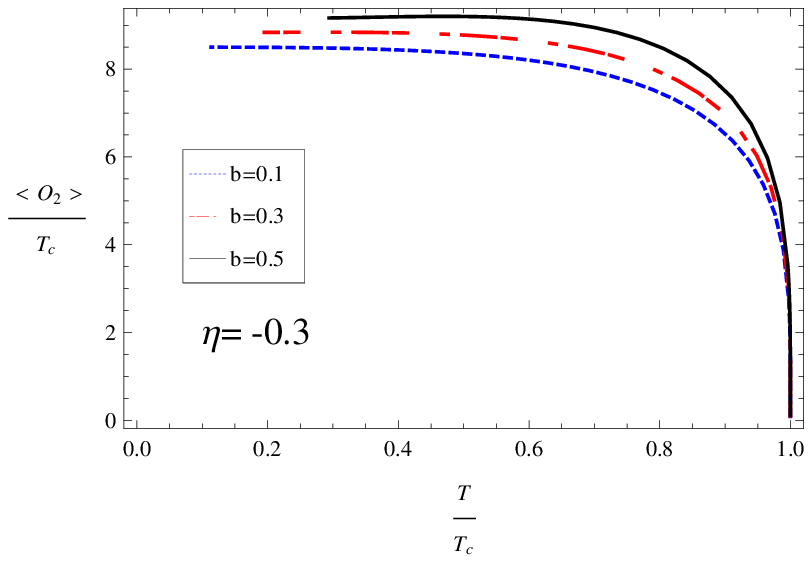}\includegraphics[width=4cm]{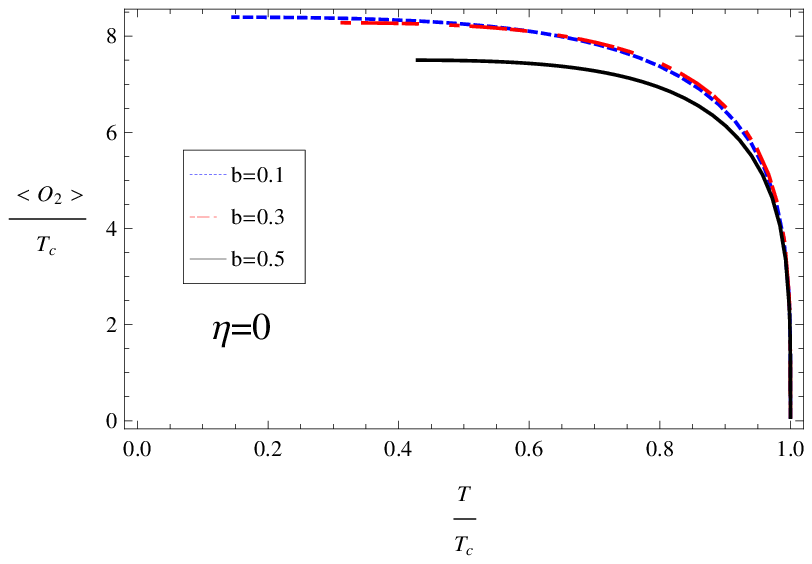}\includegraphics[width=4cm]{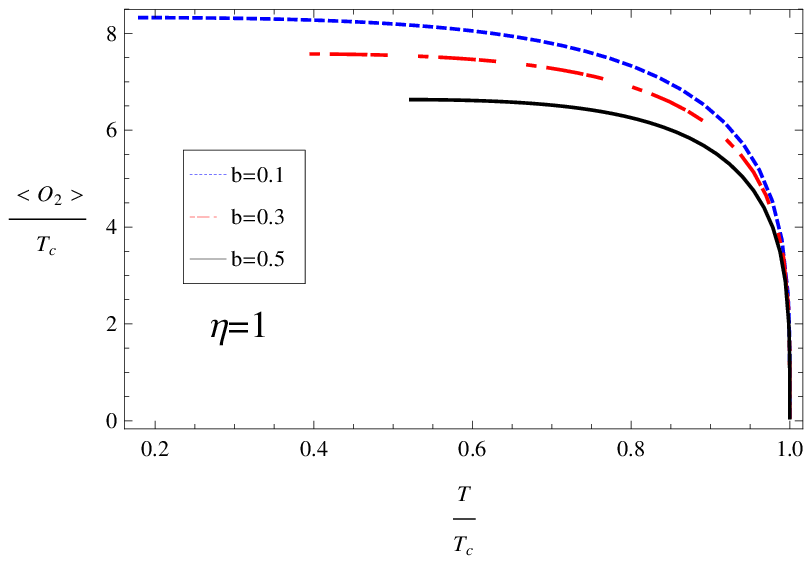}\includegraphics[width=4cm]{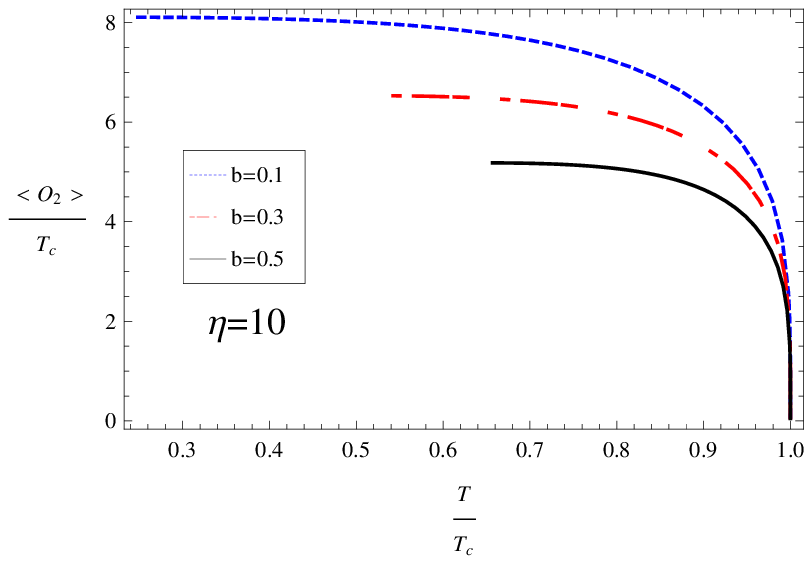}
\includegraphics[width=4cm]{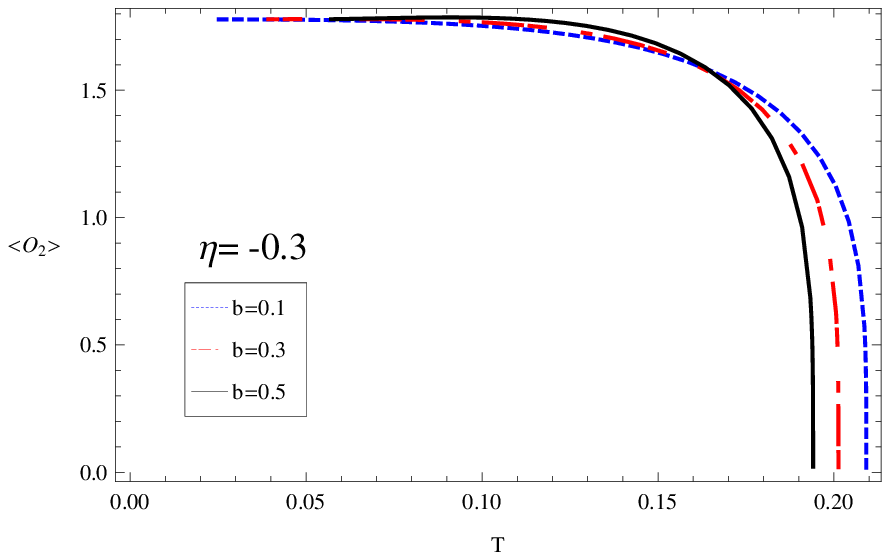}\includegraphics[width=4cm]{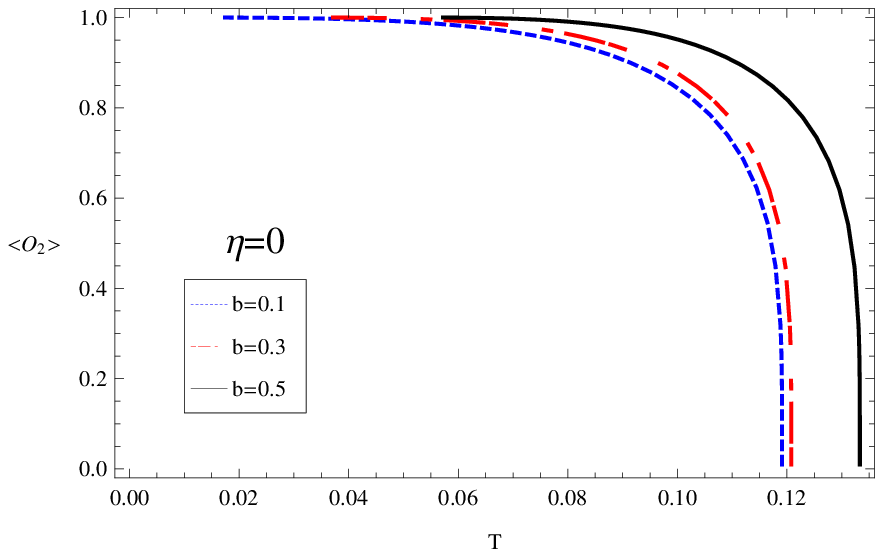}\includegraphics[width=4cm]{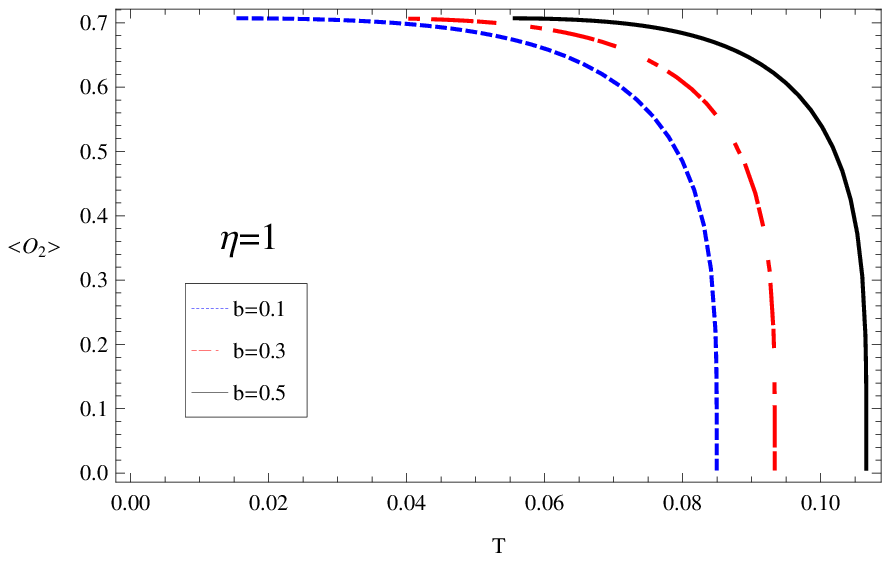}\includegraphics[width=4cm]{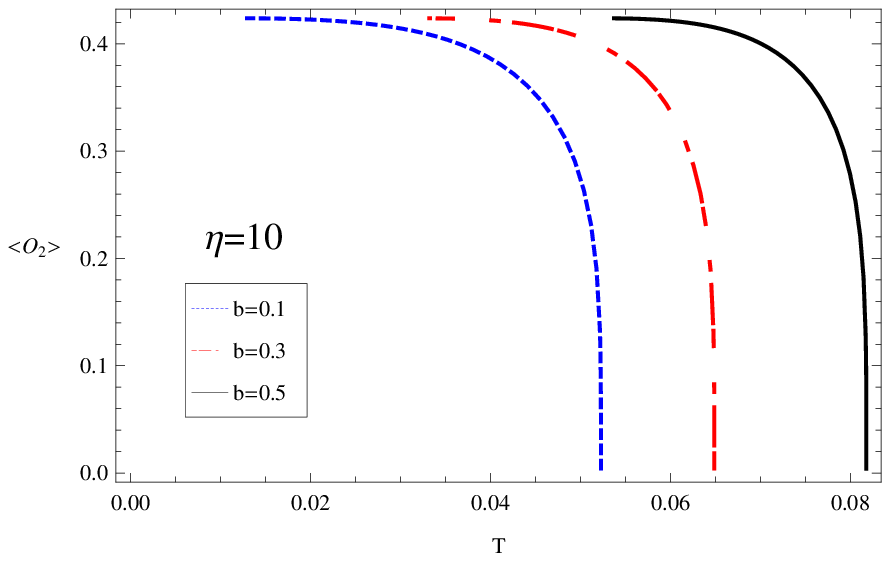}
\caption{Diagrams of phase transition for both operators $\mathcal{O}_{1,2}$ for $\eta=-0.3,0,1,10$ and $b=0.1,0.3,0.5$. In each column is shown the behaviour for both operators keeping $\eta$ fixed. } \label{fig1}
\end{figure*}

In Fig.(\ref{fig1}) we present, in each column, the diagrams of phase transition for both operators $\mathcal{O}_{1,2}$ as function of the temperature $T$ and of the dimensionless quantity $T/T_c$ keeping $\eta$ fixed and varying $b$. When $\eta<0$ the condensation becomes harder if $b$ increases and the $T_c^{SchAdS}$ for Schwarzschild AdS is larger than $T_c^{RBH}$ for the regular black hole. On the other hand, when $\eta\geq0$ the condensation becomes easier if $b$ increases and the $T_c^{RBH}>T_c^{SchAdS}$. However, in the latter case the critical temperature is smaller than first one. For small temperatures the behaviour of condensates are different with $\mathcal{O}_2$ being finite and $\mathcal{O}_1$ diverging. Thus, the regularity in the origin, characterized by the parameter $b$, has an important influence in the condensate once it can increase the critical temperature if $\eta>0$.

\begin{table}[h]
\caption{Critical parameters for $\mathcal{O}_1$ and $\mathcal{O}_2$}
\begin{center}
\begin{tabular}{|c|c|c|c|c|c|c|c|c|c|c|}
  \hline
  $\mathcal{O}_1$&$\eta$& 10 & 8 & 6 & 4 & 2 & 0 & -1/6 & -1/5 & -1/30 \\ \hline
  &$b_c$ &0.5652 & 0.5662 & 0.5678 & 0.5709 & 0.5795? & 0.7379 & 1.1984 & 1.8382 & 10.1792 \\
 \hline
 $\mathcal{O}_2$&  $\eta$& 10 & 8 & 6 & 4 & 2 & 0 & -1/30 & -1/15 & -11/100 \\ \hline
  &$b_c$ &1.562 & 1.570? & 1.583 & 1.609 & 1.685 & 4.301 & 5.296 & 7.346 & 20.154 \\
  \hline
\end{tabular}
\end{center}
\label{tab2}
\end{table}
Now keeping $b$ fixed and varying $\eta$ one can see the influence of the non-minimal coupling in the condensate. In Fig(\ref{fig2}) is shown, in each column, the diagrams of phase transition for both operators. Comparing the different curves for the same $b$ one can see that the condensation becomes harder if $\eta$ increases. When $b$ increases all the curves get closer to the critical curve for $\eta_0 =-1/3$.
\begin{figure*}[h]
\begin{center}
\includegraphics[width=5cm]{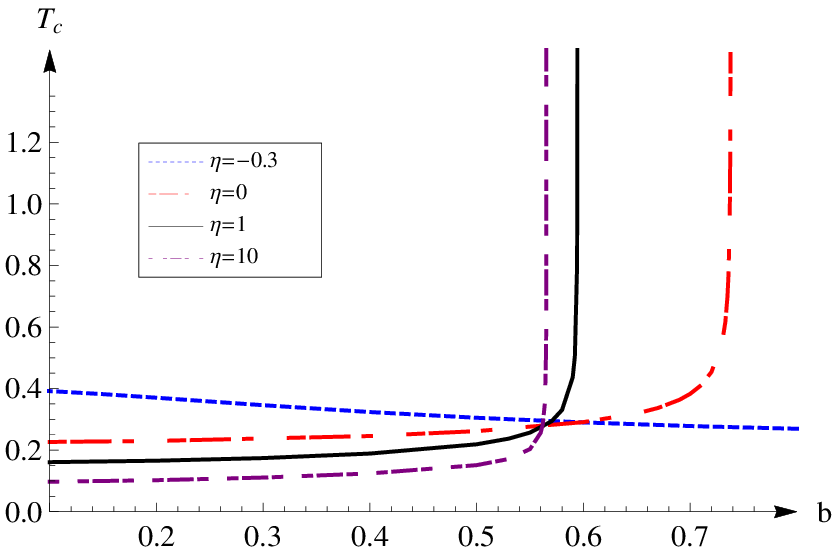}\includegraphics[width=5cm]{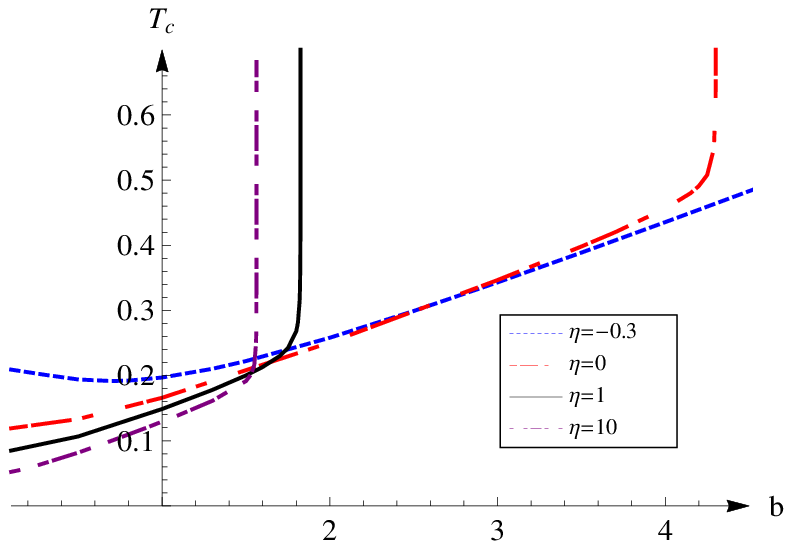}\\
\includegraphics[width=5cm]{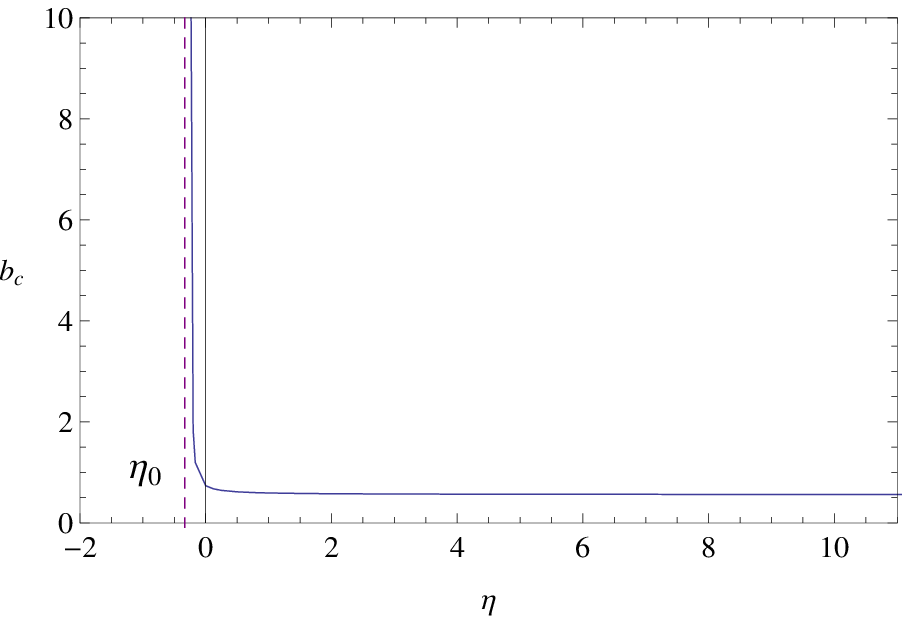}\includegraphics[width=5cm]{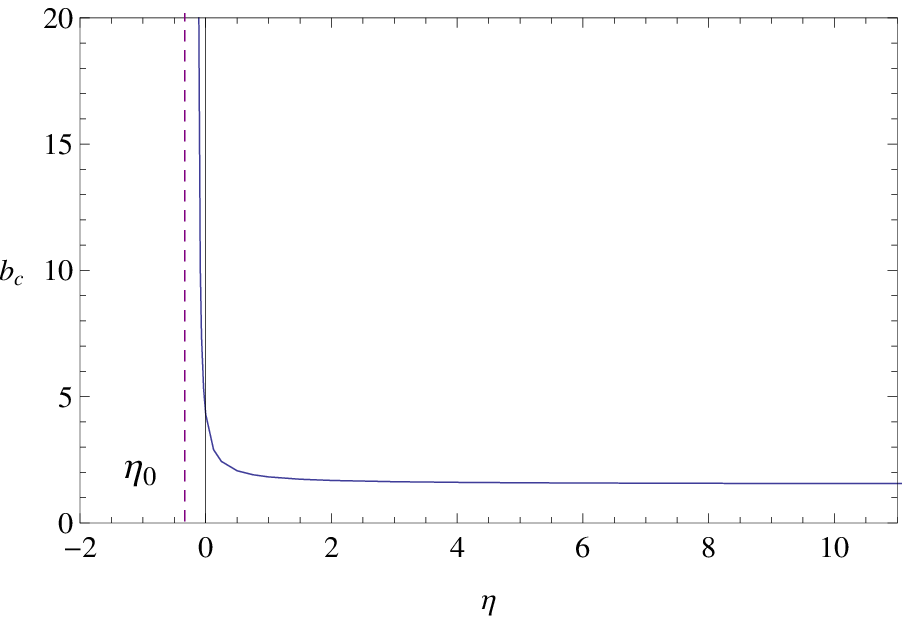}
\caption{The critical temperature $T_c$ as function of $b$ (above) and the critical parameter $b_c$ as function of $\eta$ (bellow) for the both operators ${\cal O}_1$ (left side) and ${\cal O}_2$ (right side).}
\label{fig5}
\end{center}
\end{figure*}
Despite  that behaviour, we found a critical $b_c$ in which the critical temperature increases without bound. The relation of $T_c$, $\eta$ and $b_c$ is shown in Figs.(\ref{fig5}). The critical values for each operator are compiled on the Table \ref{tab2}. One can see that near to $b_c$ the critical temperature $T_c$ increases rapidly. Moreover, $b_c$ decreases when $\eta$ increases. Thus, the non-minimal coupling, characterized by $\eta$ has also an important influence in the formation of the condensate once it affects the critical temperature decreasing the value of $b_c$.
The existence of an unlimited critical temperature can be a strong evidence that high $T_c$ superconductors must be related to the absence of the singularity in the AdS/CFT context.

\begin{figure*}[h]
\includegraphics[width=4.5cm]{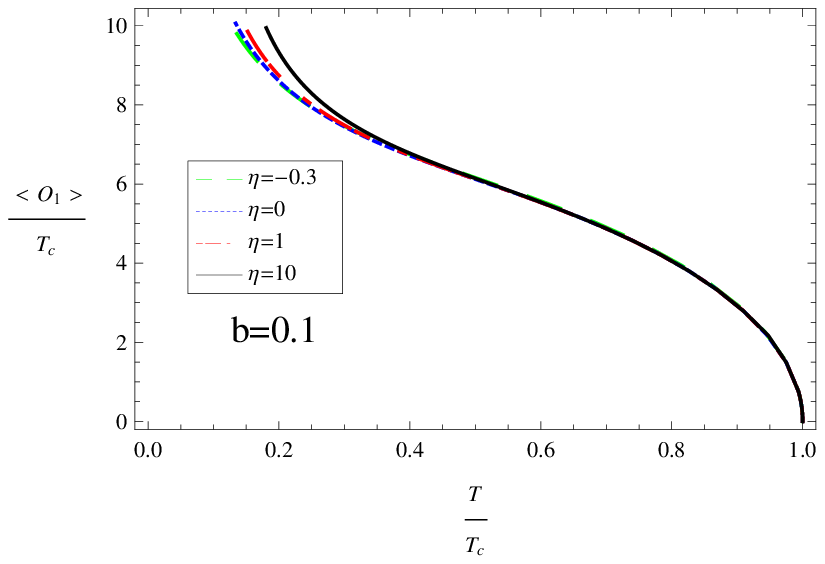}\includegraphics[width=4.5cm]{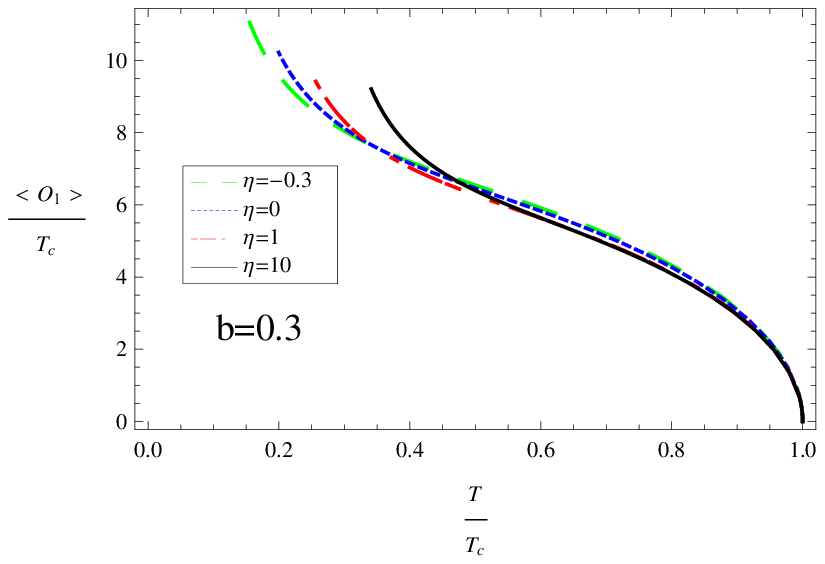}\includegraphics[width=4.5cm]{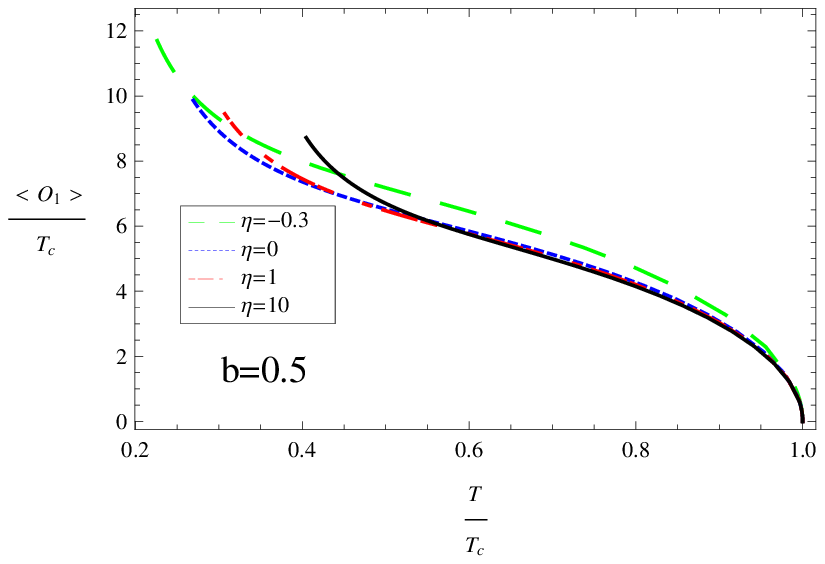}
\includegraphics[width=4.5cm]{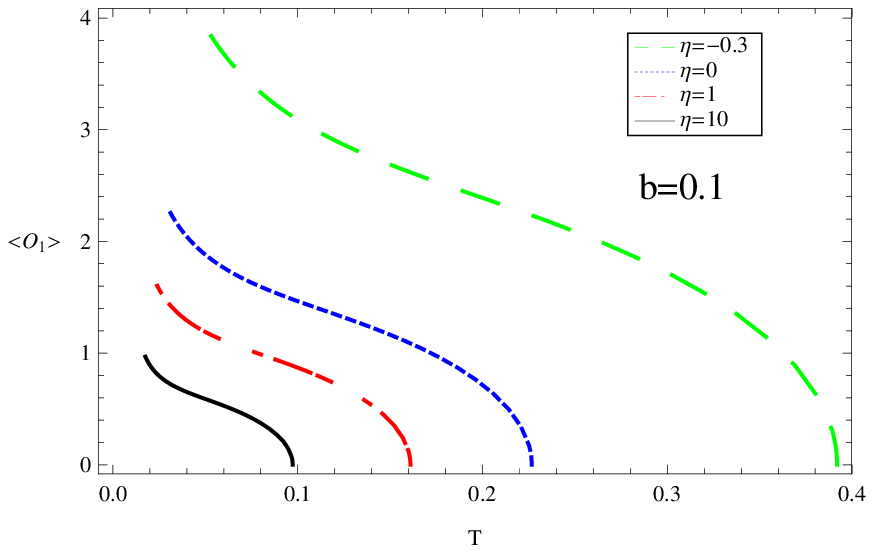}\includegraphics[width=4.5cm]{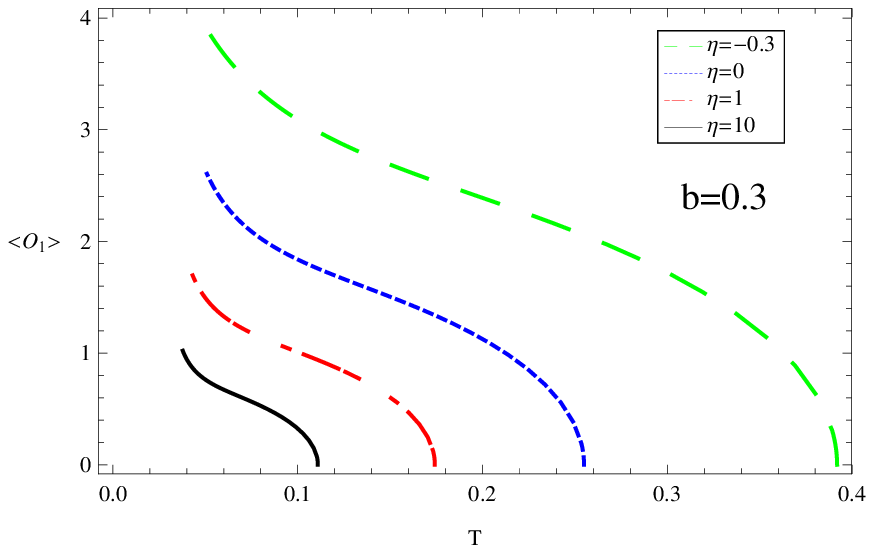}\includegraphics[width=4.5cm]{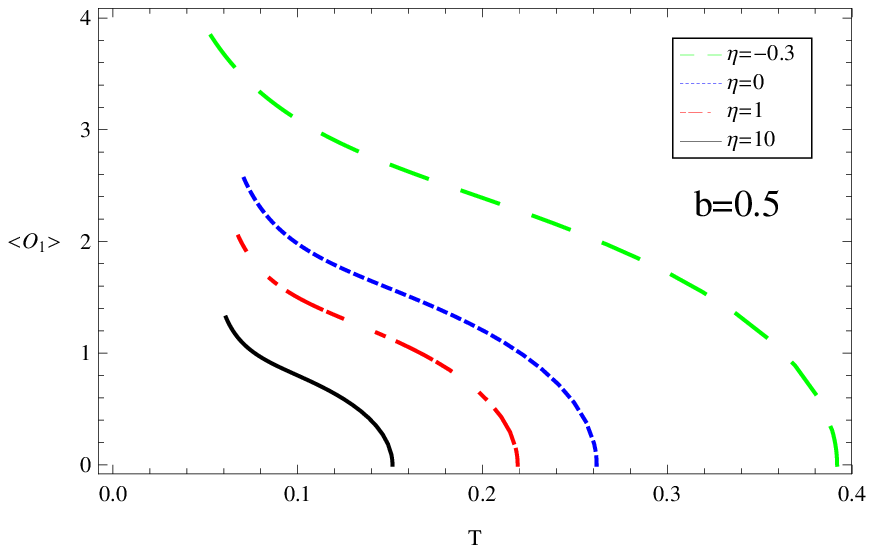}
\includegraphics[width=4.5cm]{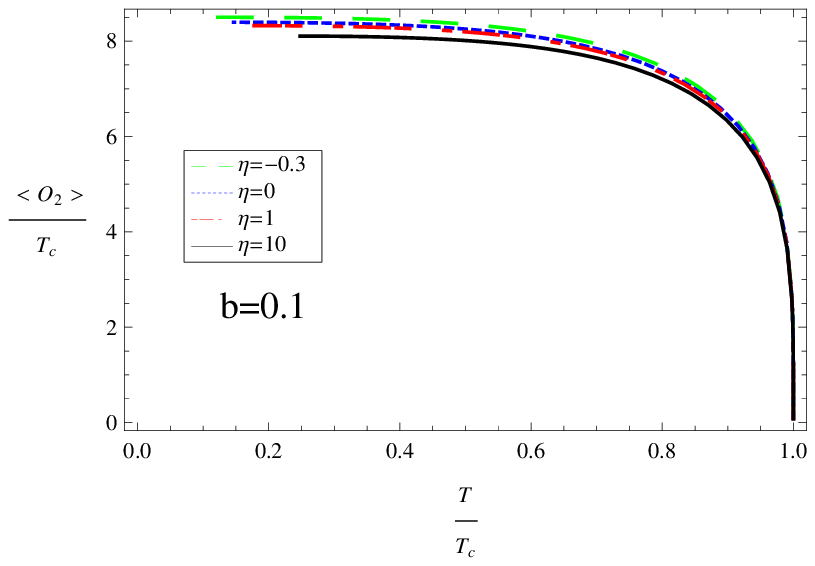}\includegraphics[width=4.5cm]{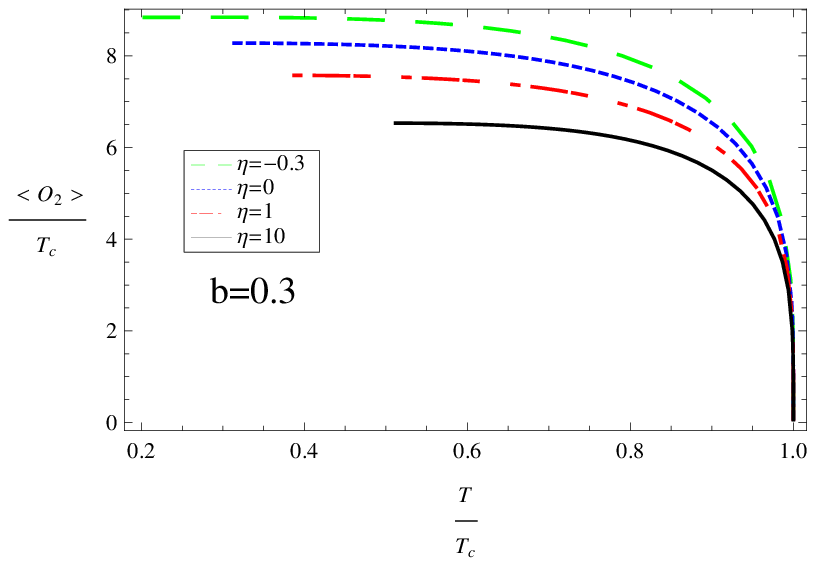}\includegraphics[width=4.5cm]{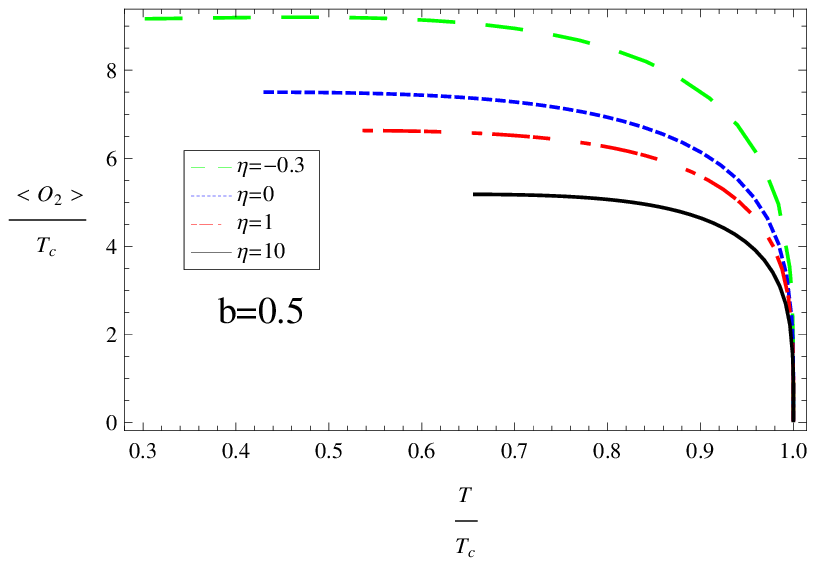}
\includegraphics[width=4.5cm]{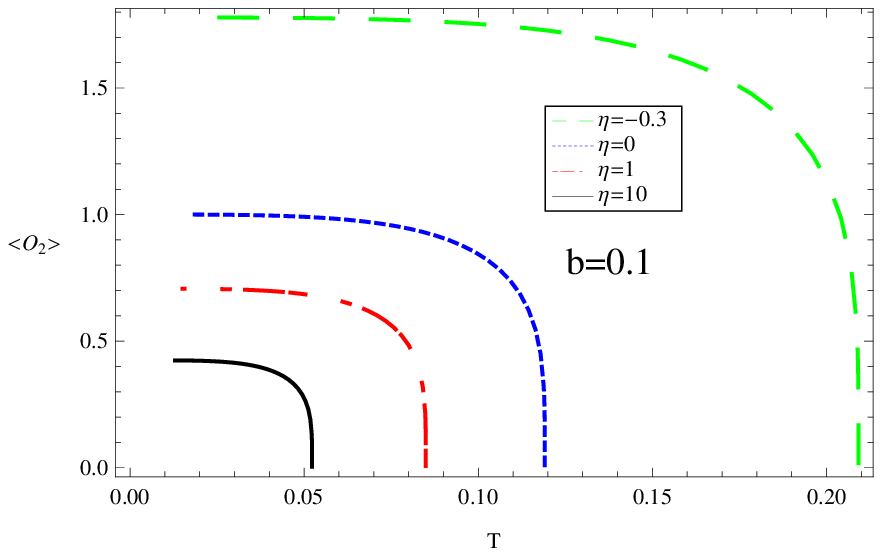}\includegraphics[width=4.5cm]{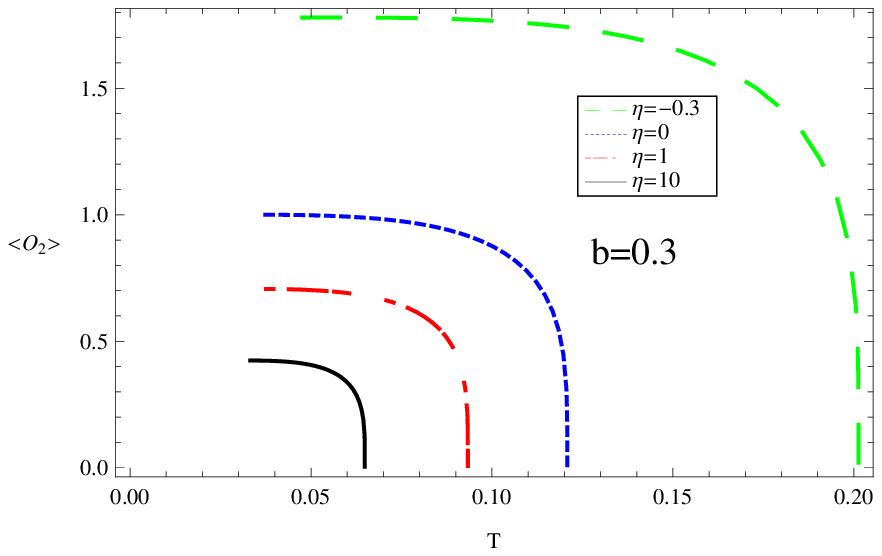}\includegraphics[width=4.5cm]{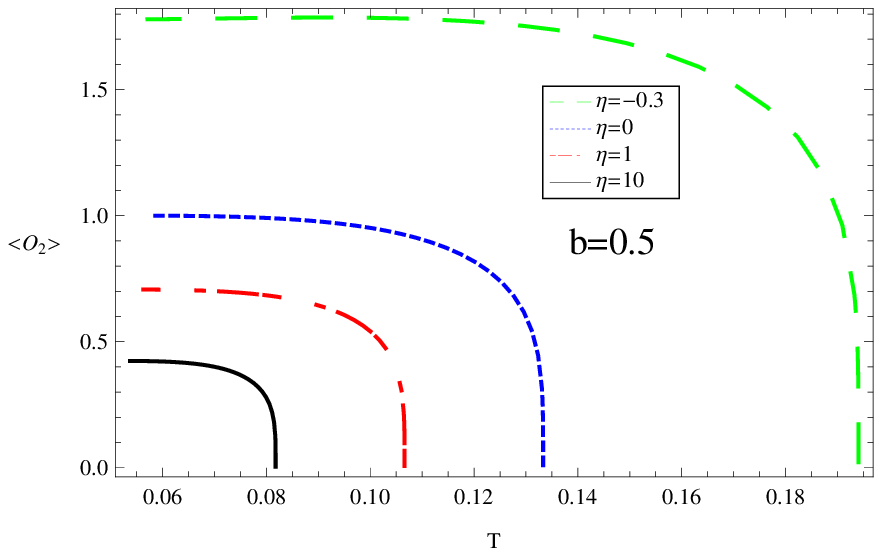}
\caption{Diagrams of phase transition for both operators $\mathcal{O}_{1,2}$ for $\eta=-0.3,0,1,10$ and $b=0.1,0.3,0.5$.In each column is shown the behaviour for both operators keeping $b$ fixed.} \label{fig2}
\end{figure*}

\section{Electrical Conductivity}

The electrical conductivity of the holographic superconductor can be computed when another component of $A_{\mu}$ is turned on. Working with a time dependent component of the Maxwell field $A_x(r,t)=A_x(r)e^{-i\omega t}$, its equation of motion can be derived, such that
\bqn
\label{eqAx}
A_x''+\frac{f'}{f}\ A_x'+\left(\frac{\omega^2}{f^2}-\frac{2q^2\Psi^2}{f}\right)A_x=0\quad ,
\eqn
The asymptotic behaviours of $A_x$ at $r_e$ and at the infinity can be obtained by the Eq.(\ref{eqAx}) and
\bqn
A_x&=&f(r)^{-\frac{i\omega}{3r_e}} \qquad\textrm{when}\ r\rightarrow r_e\\
\nb\\
\label{Axinf}
A_x&=&A_x^{(0)}+\frac{A_x^{(1)}}{r} \qquad\textrm{when}\ r\rightarrow \infty
\eqn
where we had imposed the ingoing wave boundary conditions at the event horizon. Since in the theory on the boundary the source and expectation value of the current
are related to  Eq.(\ref{Axinf}) we can rewrite it as
\bqn
\label{AxAdS}
A_x&=&A_x^{(0)}\; ,\qquad \langle J_x\rangle= A_x^{(1)}\quad .
\eqn
Using the prescription presented in \cite{holographic0} and considering the Ohm's law, we find that the conductivity in the dual theory as a function of frequency is given by
 \bqn
\label{eq9} \sigma(\omega)=-\frac{iA_x^{(1)}}{\omega A_x^{(0)}},
 \eqn

\begin{figure*}[h]
\includegraphics[width=4.5cm]{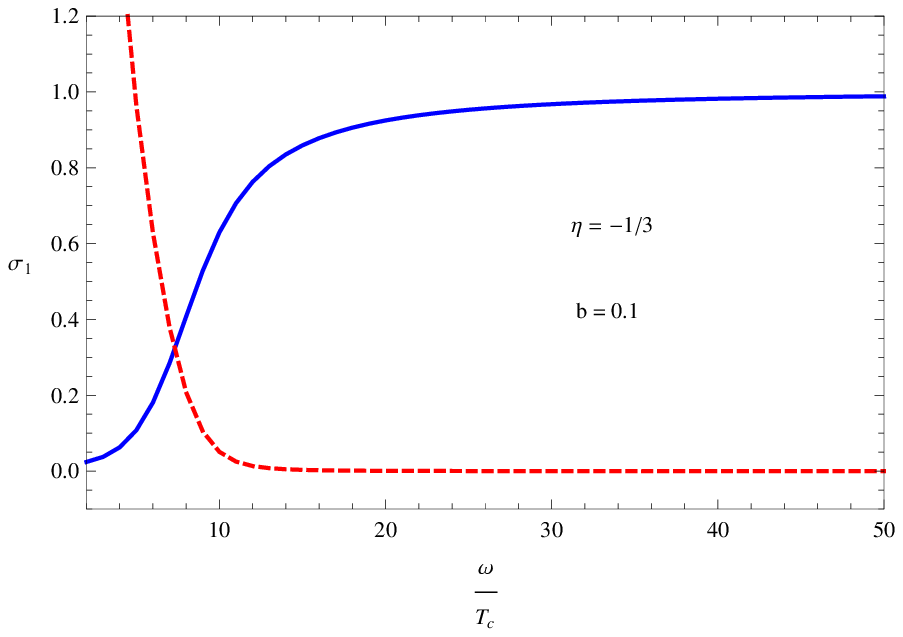}\includegraphics[width=4.5cm]{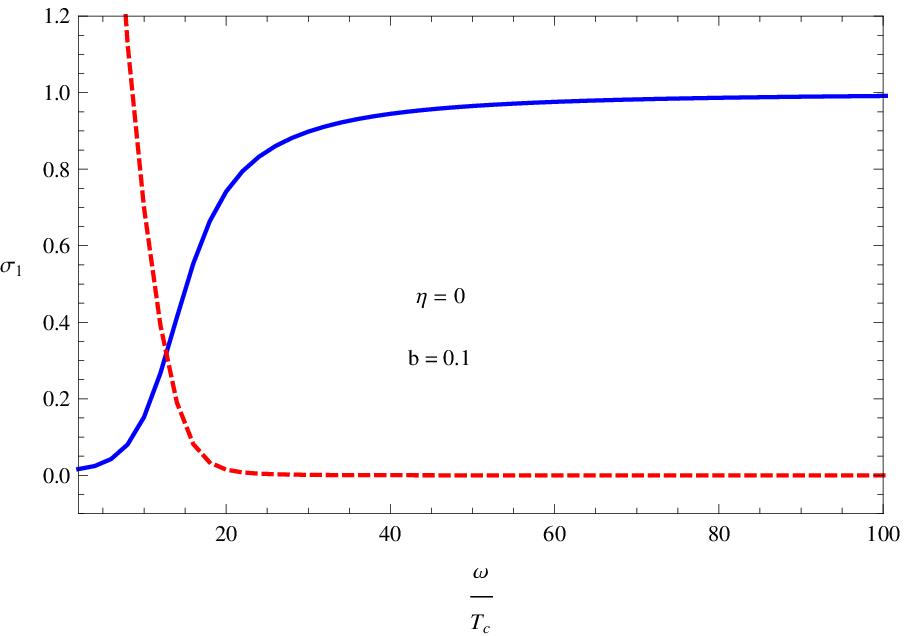}\includegraphics[width=4.5cm]{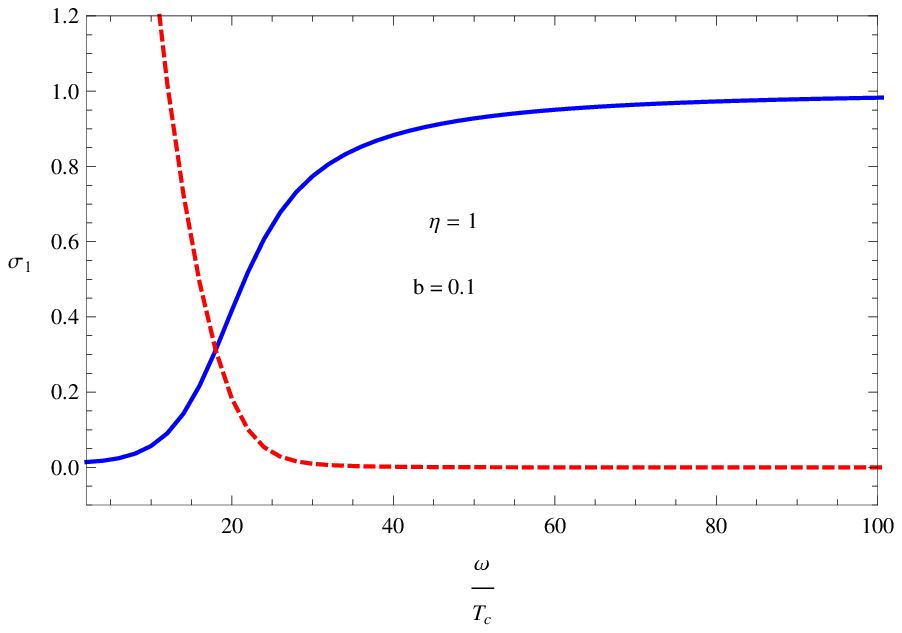}\\
\includegraphics[width=4.5cm]{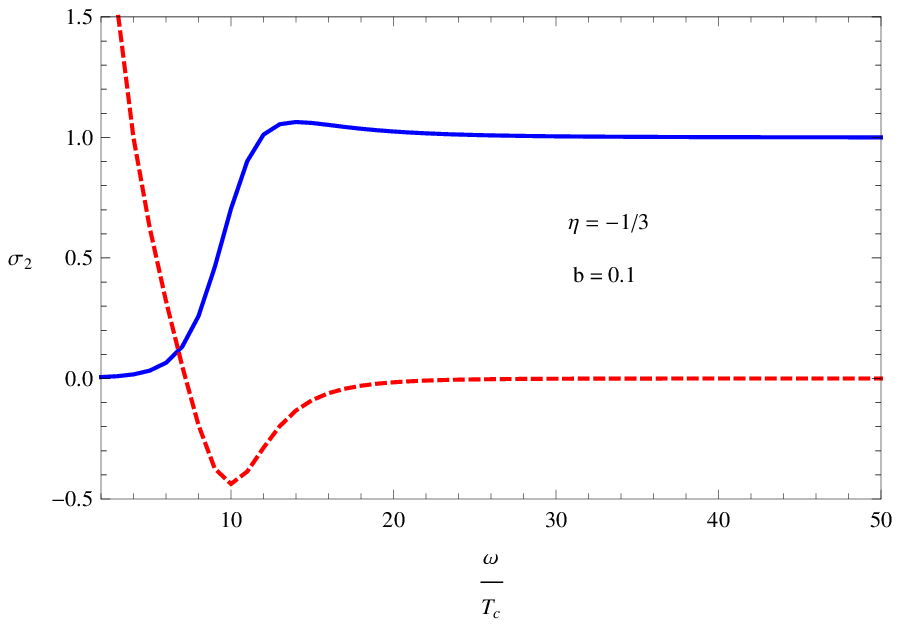}\includegraphics[width=4.5cm]{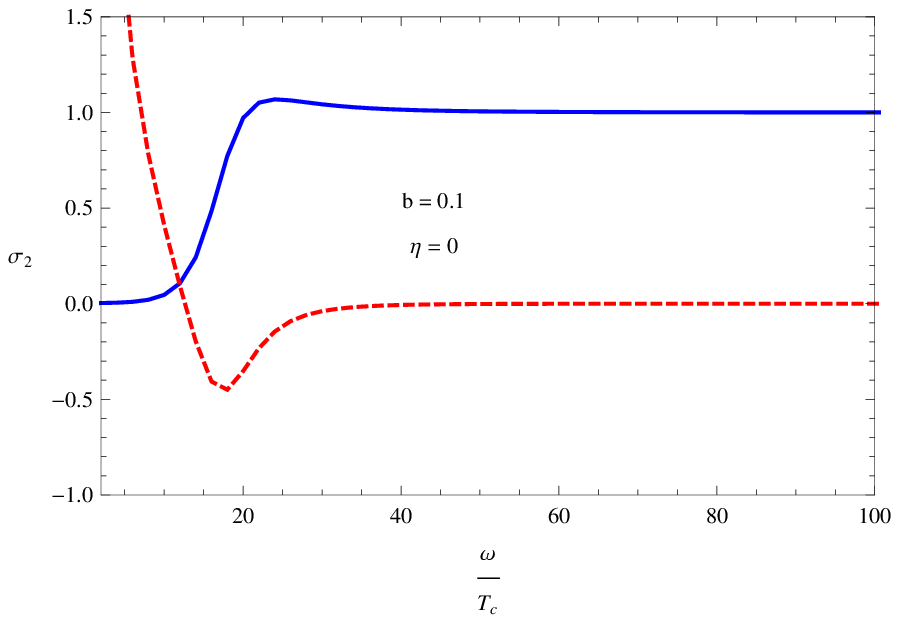}\includegraphics[width=4.5cm]{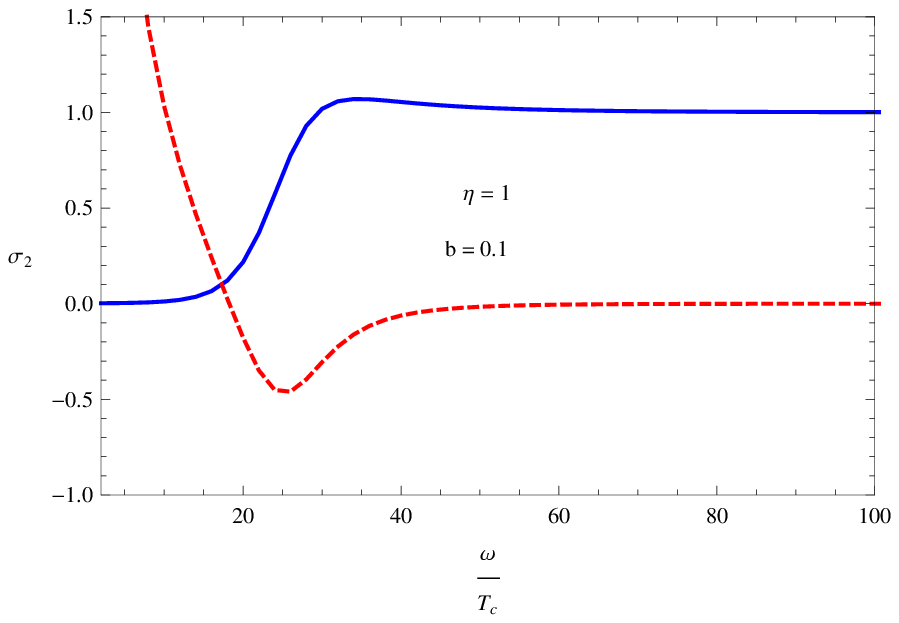}\\
\includegraphics[width=4.5cm]{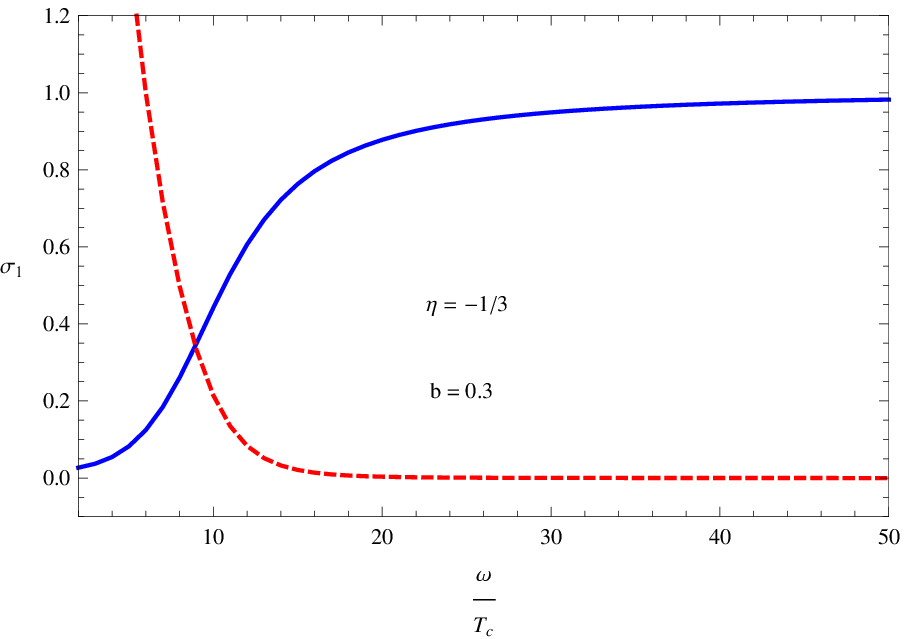}\includegraphics[width=4.5cm]{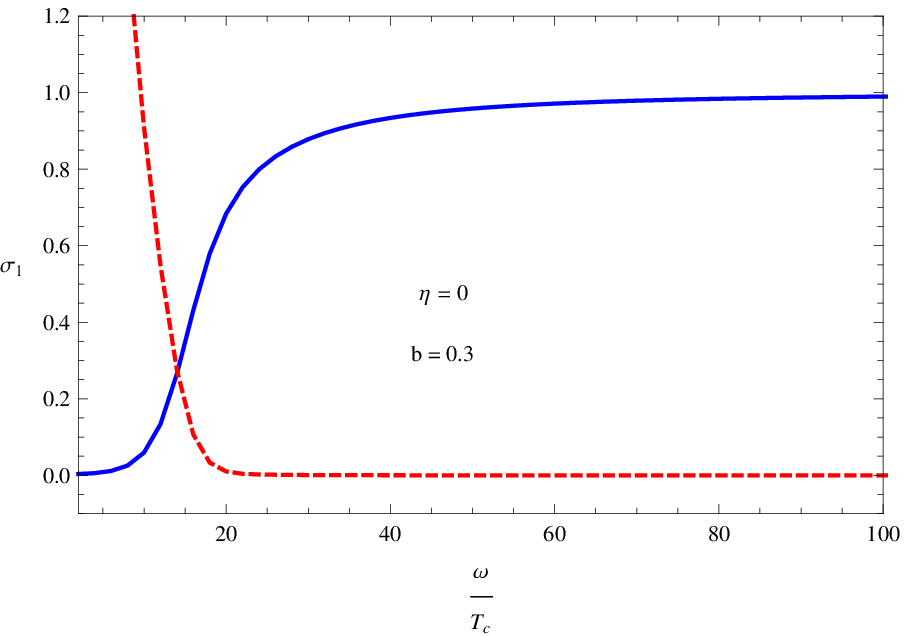}\includegraphics[width=4.5cm]{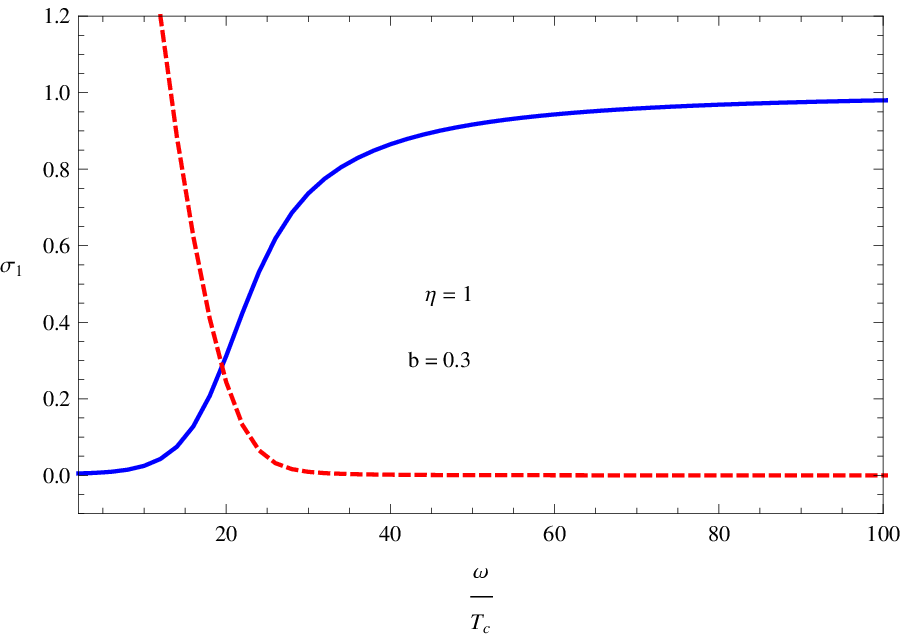}\\
\includegraphics[width=4.5cm]{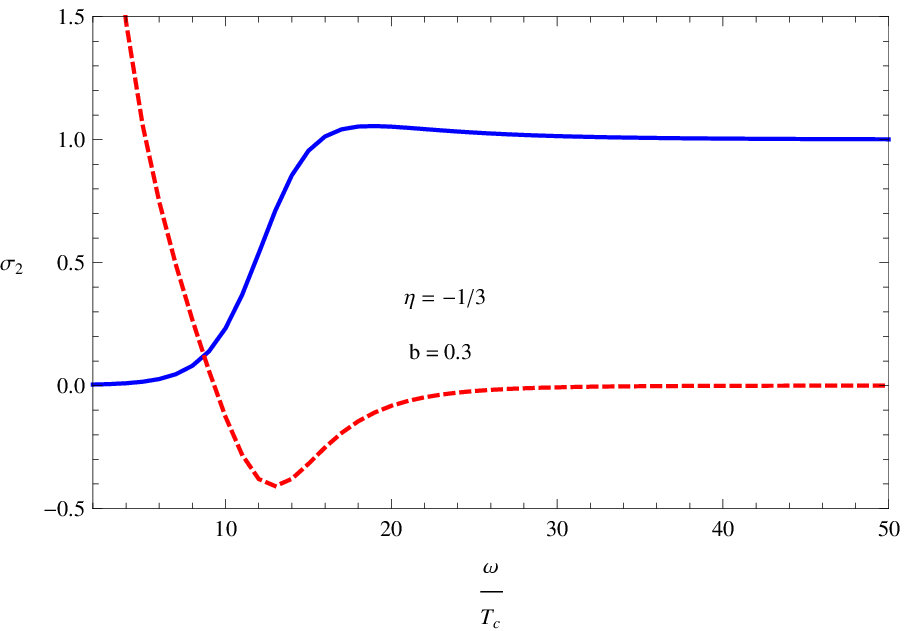}\includegraphics[width=4.5cm]{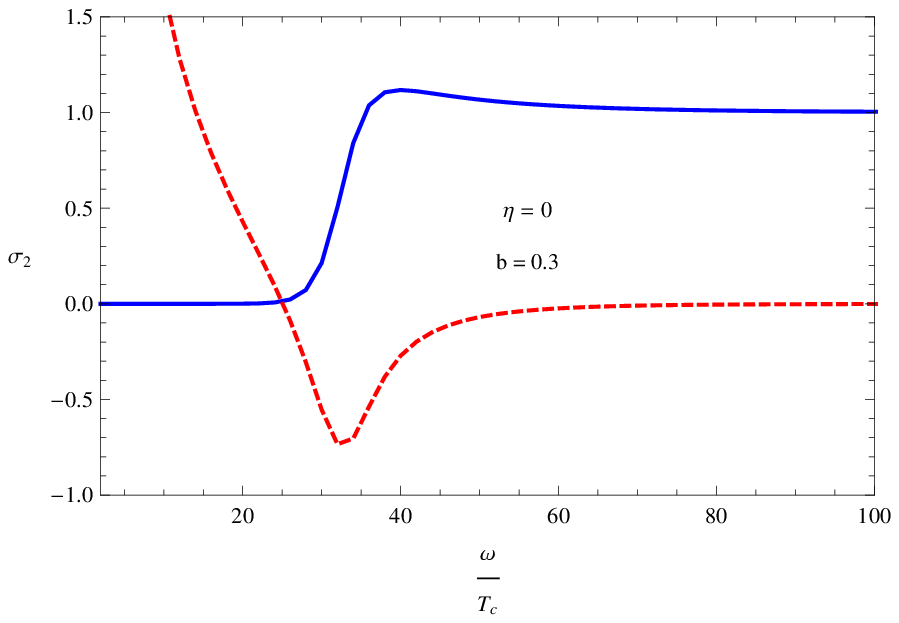}\includegraphics[width=4.5cm]{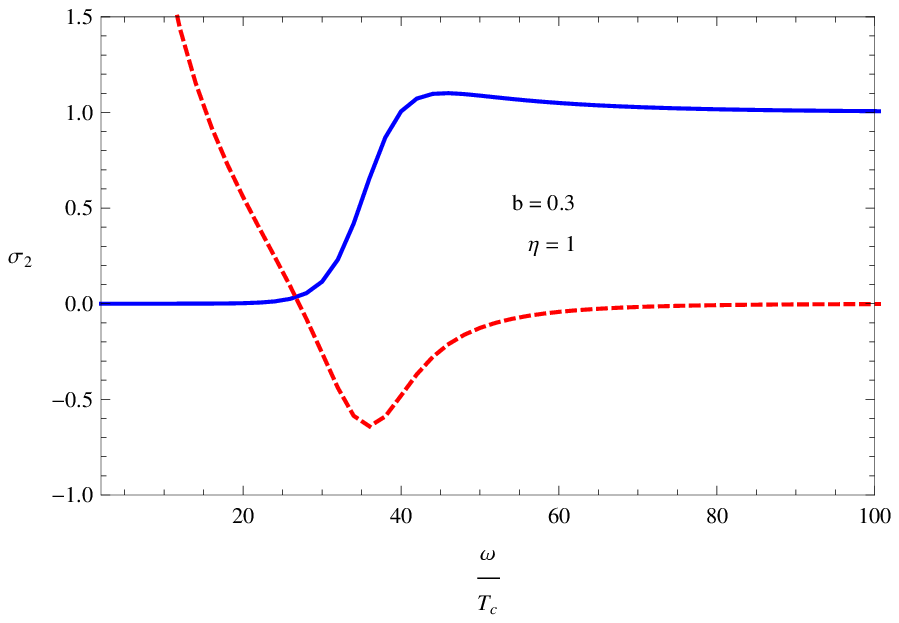}\\
\includegraphics[width=4.5cm]{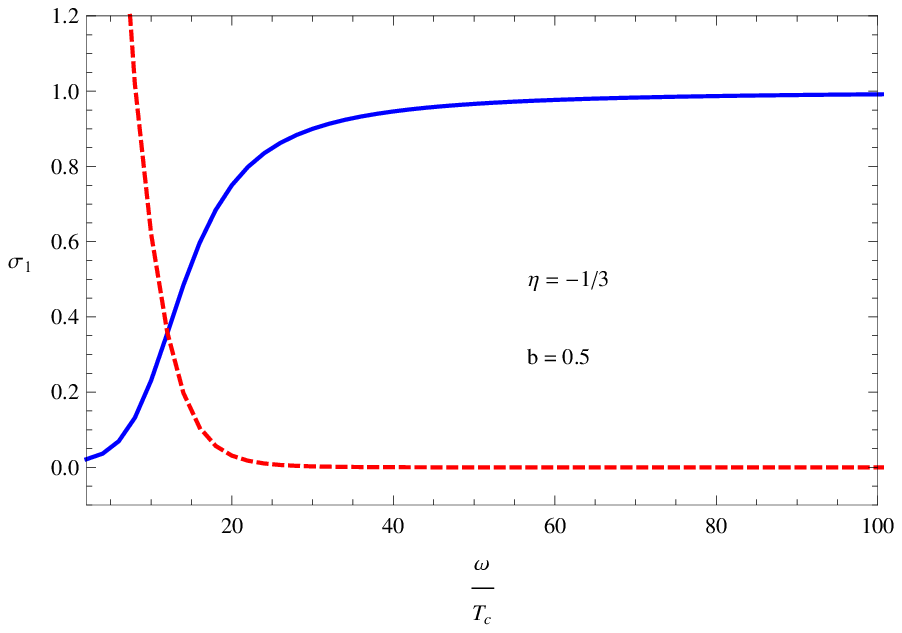}\includegraphics[width=4.5cm]{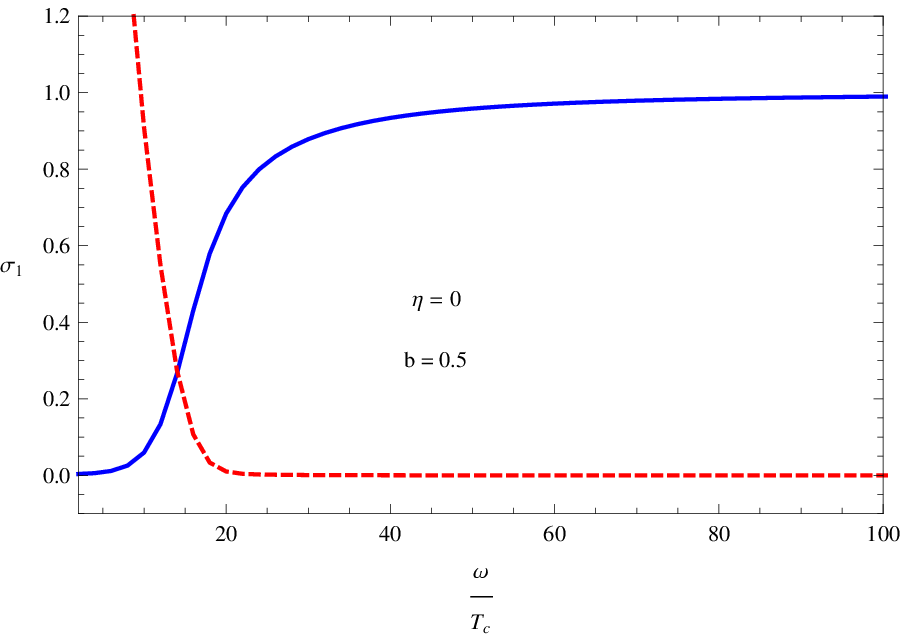}\includegraphics[width=4.5cm]{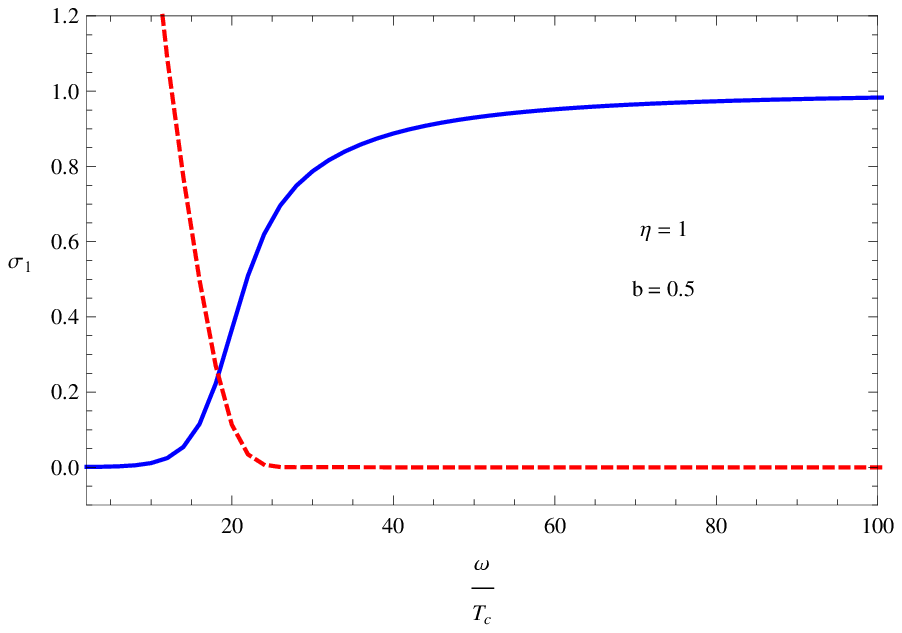}\\
\includegraphics[width=4.5cm]{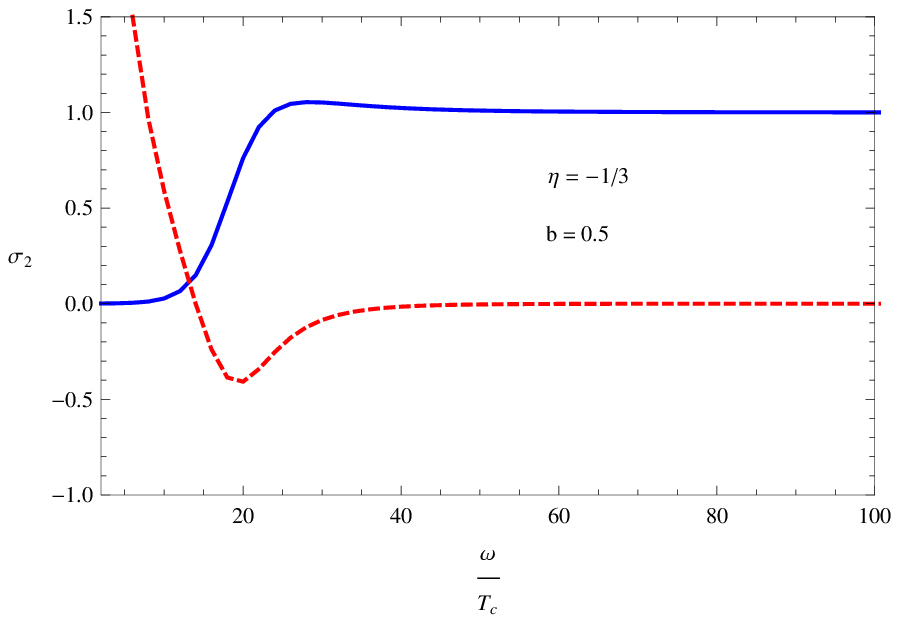}\includegraphics[width=4.5cm]{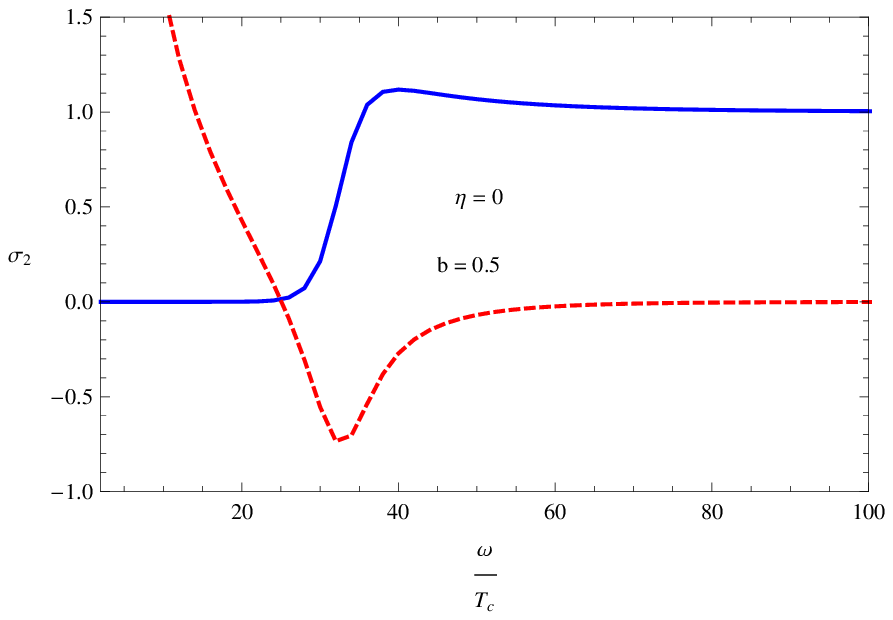}\includegraphics[width=4.5cm]{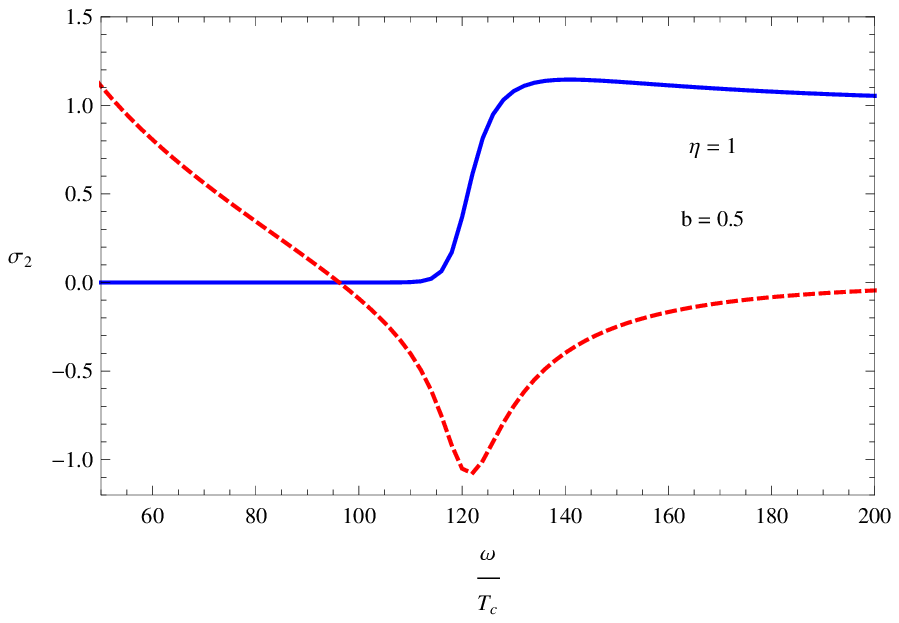}
\caption{Imaginary and real part of the electrical conductivity for both operators and different values for $b$ and $\eta$ with $T/T_c=0.5$.} \label{fig4}
\end{figure*}

The imaginary and real component of $\sigma$ are numerically calculated and are shown in  Fig.(\ref{fig4}).

It is obvious that the effects of $b$ and $\eta$ are weak in the conductivity. When $\eta$ is not negative, the curve for $\text{Re}(\sigma_i)$ and $\text{Im}(\sigma_i)$ are higher as $b$ increases, but these curves become lower if $\eta$ is negative.
Thus, we have a model sensitive to the defining couplings. In particular, we can raise the critical temperature to find superconductors at room temperature.

\section{Conclusion}

In this paper, we have derived the a new static regular phantom planar black hole, and investigated the holographic superconductor with non-minimal coupling in such spacetime. Our numerical results show that the parameter of regularity $b$ and the parameter of non-minimal coupling scalar field $\eta$ can affect the formation of the condensate and the conductivity in the superconductor. Surprisingly, we found that $b$ has a critical value $b_c$ in which the critical temperature $T_c$ increases unlimited. To our knowledge this is the first time that such propriety for unlimited $T_c$ is presented. This possibility of an unlimited critical temperature sounds like an important evidence that high $T_c$ superconductor must be related to the absence of a singularity in the bulk in the AdS/CFT context.

The main problem is thus how to connect the singularity (or non-singularity) to the defining properties of superconductors to find
very high critical temperatures.

\acknowledgments
This work is supported in part  by FAPESP No. 2012/08934-0, CNPq No. 472660/2013-6, CAPES and NNSFC No.11573022 and No.11375279

\bibliographystyle{JHEP}
\bibliography{holographicsupercondarxiv}



\end{document}